\documentstyle[aaspp4,bsf,psbox,11pt]{article}

\received{2001 November 10}
\accepted{2002 January 6}

\slugcomment{accepted for publication in ApJ}

\begin{document}

\title
{The Effect of the Hall Term on the Nonlinear Evolution of the
Magnetorotational Instability: I. Local Axisymmetric Simulations}

\author
{Takayoshi Sano and James M. Stone}
\affil
{Department of Astronomy, University of Maryland, College Park, MD
20742-2421; sano@astro.umd.edu}

\begin{abstract}
The effect of the Hall term on the evolution of the magnetorotational
instability (MRI) in weakly ionized accretion disks is investigated
using local axisymmetric simulations.  First, we show that the Hall
term has important effects on the MRI when the temperature and density
in the disk is below a few thousand K and between $10^{13}$ and
$10^{18}$~cm$^{-3}$ respectively.  Such conditions can occur in the
quiescent phase of dwarf nova disks, or in the inner part (inside 10 --
100 AU) of protoplanetary disks.  When the Hall term is important, the
properties of the MRI are dependent on the direction of the magnetic
field with respect to the angular velocity vector {\boldmath $\Omega$}.
If the disk is threaded by a uniform vertical field oriented in the
same sense as {\boldmath $\Omega$}, the axisymmetric evolution of the
MRI is an exponentially growing two-channel flow without
saturation.  When the field is oppositely directed to {\boldmath
$\Omega$}, however, small scale fluctuations prevent the nonlinear
growth of the channel flow and the MRI evolves into MHD turbulence.
These results are anticipated from the characteristics of the linear
dispersion relation.  In axisymmetry on a field with
zero-net flux, the evolution of the MRI is independent of the size of
 the Hall term relative to the inductive term.
The evolution in this case is 
determined mostly by the effect of ohmic dissipation.

\end{abstract}


\section{INTRODUCTION}

The structure and evolution of accretion disks are largely determined by
angular momentum transport processes.  One of the most promising
processes is MHD turbulence driven by the magnetorotational instability
(MRI) (Balbus \& Hawley 1991).  In ideal MHD, the growth rate of the
MRI is of the order of the orbital frequency $\Omega$, and the
characteristic wavelength of the most unstable mode is $2 \pi
v_{\rm A}/\Omega$, where $v_{\rm A}$ is the Alfv{\'e}n speed.  The
nonlinear regime of the MRI has been well studied in ideal MHD using
numerical simulations.  However, in some systems,
accretion disks are expected to be only partially ionized, and
non-ideal MHD effects, which generally suppress the growth of the
MRI, must be considered.  For example, the low temperatures of accretion
(protoplanetary) disks around young stellar objects make thermal
ionization processes ineffective, so that the abundance of charged
particles is very small, and non-ideal MHD effects are important (Gammie
1996; Stone et al. 2000).  Another example is provided by dwarf nova
systems in quiescence.  The temperature of the disk in this case can be
well below $10^4$ K and the disk is only weakly ionized, so that
non-ideal MHD effects are again important (Gammie \& Menou 1998).

There are three regimes in non-ideal MHD associated with the
relative importance of different terms in the generalized Ohm's law
(see \S 2); they are the ambipolar diffusion, ohmic dissipation,
and Hall regimes.  Which term dominates depends on the
ionization fraction and density of the gas.  Ambipolar diffusion is
most important in regions of relatively low density and high
ionization (e.g., Reg{\H o}s 1997).  The linear properties of the MRI in
the ambipolar regime
have been explored by Blaes \& Balbus (1994); they find unstable modes
exist when the collision frequency of an ion with neutrals is higher than
the orbital frequency.  The nonlinear evolution of the MRI in this
regime was examined by Hawley \& Stone (1998) using
two fluid simulations.  They found that when the coupling
between ions and neutrals is weak, the turbulence in the ionized
component of the plasma excited by the MRI does not affect the motion
of neutrals very much, thus significant angular
momentum transport requires a greater coupling between the ions and
neutrals than that required for linear instability.
Brandenburg et al. (1995) and Mac Low et al. (1995) also studied the
effect of the ambipolar diffusion in the strong coupling limit in a few
models.

Ohmic dissipation becomes important when the ionization fraction of the
gas is very low.  In this case, a linear analysis (Jin 1996; Sano \&
Miyama 1999) shows that small wavelength perturbations are damped, and
the characteristic wavelength of the MRI increases in comparison to the
ideal MHD case.  The axisymmetric 2D evolution of the MRI demonstrates
that nonlinear saturation can occur due to ohmic dissipation (Sano,
Inutsuka, \& Miyama 1998), even though the corresponding ideal MHD
cases show an ever-growing channel flow without saturation (Hawley \&
Balbus 1992).  Fleming, Stone, \& Hawley (2000) examined the nonlinear
evolution using local 3D simulations; they found that dissipation
weakens the MHD turbulence.  For significant turbulence
and angular momentum transport to occur, a critical value for the
magnetic Reynolds number must be exceeded, and this value depends
on the field geometry in the disk.

Recently, linear analyses of the MRI in the Hall regime have been
presented by Wardle (1999) and Balbus \& Terquem (2001).  The maximum
growth rate and characteristic wavelength of the MRI are strongly
modified by the Hall effect.  Most interesting is that the linear
properties of the instability depend on the direction of the magnetic
field.  This is because the dispersion relation for incompressible
Alfv{\'e}n waves traveling along field lines is quite different in Hall
MHD.  In particular, the left- and right-circularly polarized
Alfv{\'e}n waves have different phase velocities, and these two waves
interact with the Coriolis force in the disk in different ways.  One of
these two waves is commonly referred to as the whistler wave.

The purpose of this paper is to investigate the effect of the Hall term
on the nonlinear evolution of the MRI in axisymmetry.
When the Hall term is important, ohmic dissipation often cannot be
neglected (Balbus \& Terquem 2001).  Thus, we solve an induction
equation that includes both ohmic dissipation and the Hall effect.
The plan of this paper is as follows.  We examine when the Hall term
becomes important in dwarf nova and protoplanetary disks in \S 2.  Our
numerical method and the initial conditions are described in \S 3.  
The results of 2D MHD simulations are presented in \S 4 for both a
uniform and a zero-net flux vertical field.  The application of the
results to actual accretion disks is discussed in \S 5, and our results
are summarized in \S 6.

\section{THE HALL REGIME IN ACCRETION DISKS}

In this section we examine when the Hall term becomes important in
actual accretion disks.  We consider a weakly ionized fluid composed of
ions, electrons, and neutrals.  Charged dust grains can be important in
some situations (Wardle \& Ng 1999; Sano et al. 2000), because the
ionization fraction of the plasma is strongly affected by the abundance
and size distribution of the grains through recombination processes on
grain surface.  If dust grains with the
interstellar MRN size distribution (Mathis, Rumpl, \& Nordsieck 1977)
are considered, the Hall term is important for densities between
$10^7$ and $10^{11}$ cm$^{-3}$ (Wardle \& Ng 1999).  At these
densities, negatively charged grains and positively charged ions are
the dominant charge carriers.  At higher densities than $10^{11}$
cm$^{-3}$, both the negative and positive charge carriers are grains,
and the dynamics is dominated by ohmic dissipation.  The expected
density of protoplanetary disks is higher than the Hall regime in
a dusty plasma, thus if dust is present most of the disk will be in the
ohmic dissipation regime (Wardle \& Ng 1999).  However, this assumes the dust
grains are well mixed, and are unevolved.  In fact, dust grains in the
disk may grow in size through collisions and/or sediment toward the
midplane.  These evolutionary effects reduce the abundance of charged
grains, so that ions and electrons become the dominant charge carriers
even in very dense regions (Sano et al. 2000).  
On the other hand, whether the gas flow in the disk is laminar or
turbulent can largely affect the evolution of the grains (e.g., Cuzzi,
Dobrovolskis, \& Champney 1993; Hodgson \& Brandenburg 1998).
Self-consistent models of dusty disks which include the effects of
gravitational settling,
turbulent mixing, and grain growth are beyond the scope of this paper.
The analysis presented here, which does not include the effect of dust
grains, could be applicable to protoplanetary disks in the late stages
of evolution.
Dust is not expected to be important in dwarf nova disks.

Charge neutrality, $n_{e} = n_{i}$, is
assumed, where $n_{e}$ and $n_{i}$ are the number densities of
electrons and ions respectively.  The coupling between charged
particles and neutrals depends on the collision rate $\langle \sigma v
\rangle_{\alpha}$, where $\alpha$ is $i$ (ions) and $e$ (electrons).
Here, we adopt the values $\langle \sigma v \rangle_{i} = 1.9 \times
10^{-9}$ cm$^3$ s$^{-1}$ and $\langle \sigma v \rangle_{e} = 8.28
\times 10^{-10} T^{1/2}$ cm$^3$ s$^{-1}$ (Draine, Roberge, \& Dalgarno
1983).  The imperfect coupling between electrons and neutrals results
in a finite electrical conductivity,
\begin{equation}
\sigma_c = \frac{e^2 n_e}{m_e \nu_e } ~,
\end{equation}
where $e$ and $m_e$ are the electron charge and mass. 
The electron-neutral collision frequency is $\nu_e = n_n \langle \sigma
v \rangle_e$, where $n_n$ is the number density of neutrals.
The associated magnetic diffusivity is 
\begin{equation}
\eta = \frac{c^2}{4 \pi \sigma_c} ~,
\end{equation}
where c is the speed of light.
The cyclotron frequencies of the ions and electrons are 
\begin{equation}
\omega_{c \alpha} = \frac{eB}{m_{\alpha}c} ~,
\end{equation}
where $\alpha$ is $i$ and $e$.

The induction equation including the terms of the ohmic dissipation, the 
Hall effect, and the ambipolar diffusion is given by
\begin{equation}
\frac{\partial \mbox{\boldmath $B$}}{\partial t} =
\mbox{\boldmath $\nabla$} \times
\left[
\mbox{\boldmath $v$} \times \mbox{\boldmath $B$}
- \frac{4 \pi  \eta \mbox{\boldmath $J$}}{c}
- \frac{\mbox{\boldmath $J$} \times \mbox{\boldmath $B$}}{e n_e}
+ \frac{\left(\mbox{\boldmath $J$} \times \mbox{\boldmath $B$}\right)
\times \mbox{\boldmath $B$}}{c \gamma \rho_i \rho}
\right] ~,
\label{eqn:faraday}
\end{equation}
where {\boldmath $v$} is the neutral velocity and 
\begin{equation}
 \mbox{\boldmath $J$} = \frac{c}{4 \pi} 
\left( \nabla \times \mbox{\boldmath $B$} \right)
\end{equation}
is the current density.
The drag coefficient is 
\begin{equation}
\gamma = 
\frac{\langle \sigma v \rangle_{i}}{m_i + m_n} ~,
\end{equation}
where $m_i$ and $m_n = \mu m_{\rm H}$ are the ion and neutral particle mass,
$\mu$ is the mean molecular weight of the gas, and $m_{\rm H}$ is the
mass of a hydrogen atom.

Here we estimate the relative magnitude of the four terms on the right
hand side of the induction equation.  Following Balbus \& Terquem
(2001), we denote these terms (reading from left to right) as $I$
(Inductive), $O$ (Ohmic), $H$ (Hall), and $A$ (Ambipolar).  The ratios
$A/H$ and $H/O$ are given by
\begin{equation}
\frac{A}{H} = \frac{w_{ci}}{\gamma \rho} ~,
\label{eqn:ah}
\end{equation}
and
\begin{equation}
\frac{H}{O} = \frac{w_{ce}}{\nu_{e}} ~.
\label{eqn:ho}
\end{equation}
Note that these ratios are determined only by micro-physical
quantities.  In this paper we define the Hall regime as the conditions
under which the Hall term is the largest of the three non-ideal MHD
effects.  From equations (\ref{eqn:ah}) and (\ref{eqn:ho}), in the Hall
regime electrons are coupled to the magnetic field ($w_{ce} / \nu_{e} >
1$), but ions are not ($w_{ci} / \gamma \rho < 1$) because of the 
collisions with neutrals.  In low density
regions, both ions and electrons are coupled to the magnetic field:
this is the ambipolar diffusion regime ($A > H > O$).  On the other
hand, in the high density regions, both the electrons and ions are
decoupled from the field, leading to the ohmic dissipation regime ($O >
H > A$).  The density range of the Hall regime is therefore
intermediate between these two (see below).

The importance of each non-ideal MHD effect relative to the inductive
term is 
\begin{equation}
\frac{O}{I} 
= \frac{\eta}{V L}
= \frac{\eta \Omega}{v_{\rm A}^2}
\equiv Re_{M}^{-1} ~,
\label{eqn:rem}
\end{equation}
\begin{equation}
\frac{H}{I} = \frac{cB}{4 \pi e n_e V L} 
= \frac{cB\Omega}{4 \pi e n_e v_{\rm A}^2} \equiv \frac{X}2 ~,
\label{eqn:x}
\end{equation}
and
\begin{equation}
\frac{A}{I} = \frac{B^2}{4 \pi \gamma \rho_i \rho V L} 
= \frac{\Omega}{\gamma \rho_i}
\label{eqn:aoi}
\end{equation}
for the ohmic dissipation, the Hall effect, and the ambipolar
diffusion, respectively, where $V$ and $L$ are typical values for the
velocity and length scale in the fluid.  We assume that $V$ is of the
order of $v_{\rm A}$, and that $L = v_{\rm A} / \Omega$, i.e., the
characteristic scale of the MRI.  Equations (\ref{eqn:rem}) and
(\ref{eqn:x}) give the definition of two important non-dimensional
parameters; the magnetic Reynolds number $Re_{M}$ and the Hall
parameter $X$.  The relation between these two parameters is given by
\begin{equation}
\frac{H}{O} = \frac{w_{ce}}{\nu_e} = \frac{X Re_{M}}{2} ~.
\end{equation}
This is independent of the ionization fraction and 
the choice of the typical scales $V$ and $L$,
and constrains the values of $X$ and $Re_{M}$ in the Hall
regime.  Note that the definition of the Hall parameter $X$ is
the same as $x$ in Balbus \& Terquem (2001) (in this paper we shall use
$x$ to denote one of the coordinates in our computational domain).  
The importance of the Hall effect is estimated by the ratio
$H/I$, or $X$.  In order to evaluate this ratio in actual disks, we
need to know the number density of electrons $n_{e}$.  In the following
subsections, we calculate the ionization fraction, $n_e / n_n$,
in dwarf nova and protoplanetary disks.

\subsection{Dwarf Nova Disks in Queiscence}

In outburst, the typical temperature of dwarf nova disks is over $10^4$ K
so that the gas is fully ionized (e.g., Cannizzo 1993). 
However, in quiescence the temperature is only a few thousand K.
The main source of free electrons at this temperature is thermal
ionization of Na, Al, and K.
Assuming the mean mass of ions is $m_i = 30 m_{\rm H}$, and the
neutrals are hydrogen atoms ($\mu = 1.27$), equations (\ref{eqn:ah}) --
(\ref{eqn:rem}) give
\begin{equation}
\frac{A}{H} = \left( \frac{3.0 \times 10^{13}}{n_n} \right)^{1/2}
\left( \frac{T}{10^3 ~{\rm K}}\right)^{1/2}
\left( \frac{c_s}{v_{\rm A}}\right)^{-1} ~,
\label{eqn:ah2}
\end{equation}
\begin{equation}
\frac{H}{O} = \left( \frac{7.8 \times 10^{17}}{n_n} \right)^{1/2}
\left( \frac{c_s}{v_{\rm A}}\right)^{-1} ~,
\label{eqn:ho2}
\end{equation}
and
\begin{equation}
\frac{O}{I} = 1.1 \times 10^{-9} 
\left( \frac{n_e}{n_n} \right)^{-1}
\left( \frac{T}{10^3 ~{\rm K}} \right)^{-1/2} 
\left( \frac{c_s}{v_{\rm A}} \right)^2
\left( \frac{M}{M_{\odot}} \right)^{1/2}
\left( \frac{r}{10^{10} ~{\rm cm}} \right)^{-3/2} ~,
\label{eqn:oi2}
\end{equation}
where $r$ is the distance from the central star, which has mass $M$.
For $M = M_{\odot}$, $T = 10^3$ K, and $v_{\rm A}/c_{\rm s} = 1$,
the density range of the Hall regime is $10^{13} \lesssim n_{n}
\lesssim 10^{18}$ cm$^{-3}$.
The typical density of a dwarf nova disk in quiescence is $n_{n} \sim
10^{18}$ cm$^{-3}$
(Gammie \& Menou 1998), so that the disk is in the Hall regime.
The ratio $H/O$ varies from unity to $160 (T / 10^3 ~{\rm K})^{-1/2}$ in
the Hall regime, so that 
\begin{equation}
 2 \lesssim |X| Re_{M} \lesssim 320 
\left( \frac{T}{10^3~{\rm K}} \right)^{-1/2} ~.
\label{eqn:xrm}
\end{equation}

\begin{figure}
\begin{center}
\hspace*{-45pt}
\setlength{\unitlength}{0.1bp}%
{
}%
\begin{picture}(3600,2160)(0,0)%
{
}%
\put(3037,1605){\makebox(0,0)[r]{\large{${A}/{I}$}}}%
\put(3037,1762){\makebox(0,0)[r]{\large{${O}/{I}$}}}%
\put(3037,1919){\makebox(0,0)[r]{\large{${H}/{I}$}}}%
\put(1106,1840){\makebox(0,0)[l]{\large{(a) $T$ = 3000 K}}}%
\put(2200,50){\makebox(0,0){\large $\log n_{n} ~[{\rm cm}^{-3}]$}}%
\put(700,1180){%
\makebox(0,0)[b]{\shortstack{\large{$\log$ [ {\bsf Ratios to the Inductive Term} ]}}}%
}%
\put(3450,200){\makebox(0,0)[c]{\large $20$}}%
\put(3138,200){\makebox(0,0)[c]{\large $19$}}%
\put(2825,200){\makebox(0,0)[c]{\large $18$}}%
\put(2513,200){\makebox(0,0)[c]{\large $17$}}%
\put(2200,200){\makebox(0,0)[c]{\large $16$}}%
\put(1888,200){\makebox(0,0)[c]{\large $15$}}%
\put(1575,200){\makebox(0,0)[c]{\large $14$}}%
\put(1263,200){\makebox(0,0)[c]{\large $13$}}%
\put(950,200){\makebox(0,0)[c]{\large $12$}}%
\put(900,2060){\makebox(0,0)[r]{\large $2$}}%
\put(900,1840){\makebox(0,0)[r]{\large $1$}}%
\put(900,1620){\makebox(0,0)[r]{\large $0$}}%
\put(900,1400){\makebox(0,0)[r]{\large $-1$}}%
\put(900,1180){\makebox(0,0)[r]{\large $-2$}}%
\put(900,960){\makebox(0,0)[r]{\large $-3$}}%
\put(900,740){\makebox(0,0)[r]{\large $-4$}}%
\put(900,520){\makebox(0,0)[r]{\large $-5$}}%
\put(900,300){\makebox(0,0)[r]{\large $-6$}}%
\end{picture}%
\\
\hspace*{-45pt}
\setlength{\unitlength}{0.1bp}%
{
}%
\begin{picture}(3600,2160)(0,0)%
{
}%
\put(3037,442){\makebox(0,0)[r]{\large{${A}/{I}$}}}%
\put(3037,599){\makebox(0,0)[r]{\large{${O}/{I}$}}}%
\put(3037,756){\makebox(0,0)[r]{\large{${H}/{I}$}}}%
\put(1106,520){\makebox(0,0)[l]{\large{(b) $T$ = 1500 K}}}%
\put(2200,50){\makebox(0,0){\large $\log n_{n} ~[{\rm cm}^{-3}]$}}%
\put(700,1180){%
\makebox(0,0)[b]{\shortstack{\large{$\log$ [ {\bsf Ratios to the Inductive Term} ]}}}%
}%
\put(3450,200){\makebox(0,0)[c]{\large $20$}}%
\put(3138,200){\makebox(0,0)[c]{\large $19$}}%
\put(2825,200){\makebox(0,0)[c]{\large $18$}}%
\put(2513,200){\makebox(0,0)[c]{\large $17$}}%
\put(2200,200){\makebox(0,0)[c]{\large $16$}}%
\put(1888,200){\makebox(0,0)[c]{\large $15$}}%
\put(1575,200){\makebox(0,0)[c]{\large $14$}}%
\put(1263,200){\makebox(0,0)[c]{\large $13$}}%
\put(950,200){\makebox(0,0)[c]{\large $12$}}%
\put(900,2060){\makebox(0,0)[r]{\large $2$}}%
\put(900,1840){\makebox(0,0)[r]{\large $1$}}%
\put(900,1620){\makebox(0,0)[r]{\large $0$}}%
\put(900,1400){\makebox(0,0)[r]{\large $-1$}}%
\put(900,1180){\makebox(0,0)[r]{\large $-2$}}%
\put(900,960){\makebox(0,0)[r]{\large $-3$}}%
\put(900,740){\makebox(0,0)[r]{\large $-4$}}%
\put(900,520){\makebox(0,0)[r]{\large $-5$}}%
\put(900,300){\makebox(0,0)[r]{\large $-6$}}%
\end{picture}%
\\
\caption
{Ratios of non-ideal MHD terms to the inductive term given by equations
(\protect{\ref{eqn:rem}}) through (\protect{\ref{eqn:aoi}}) for the case
of a dwarf nova disk in quiescence for (a) $T = 3000$ K and (b) $T =
1500$ K.
\label{fig:dn}}
\end{center}
\end{figure}
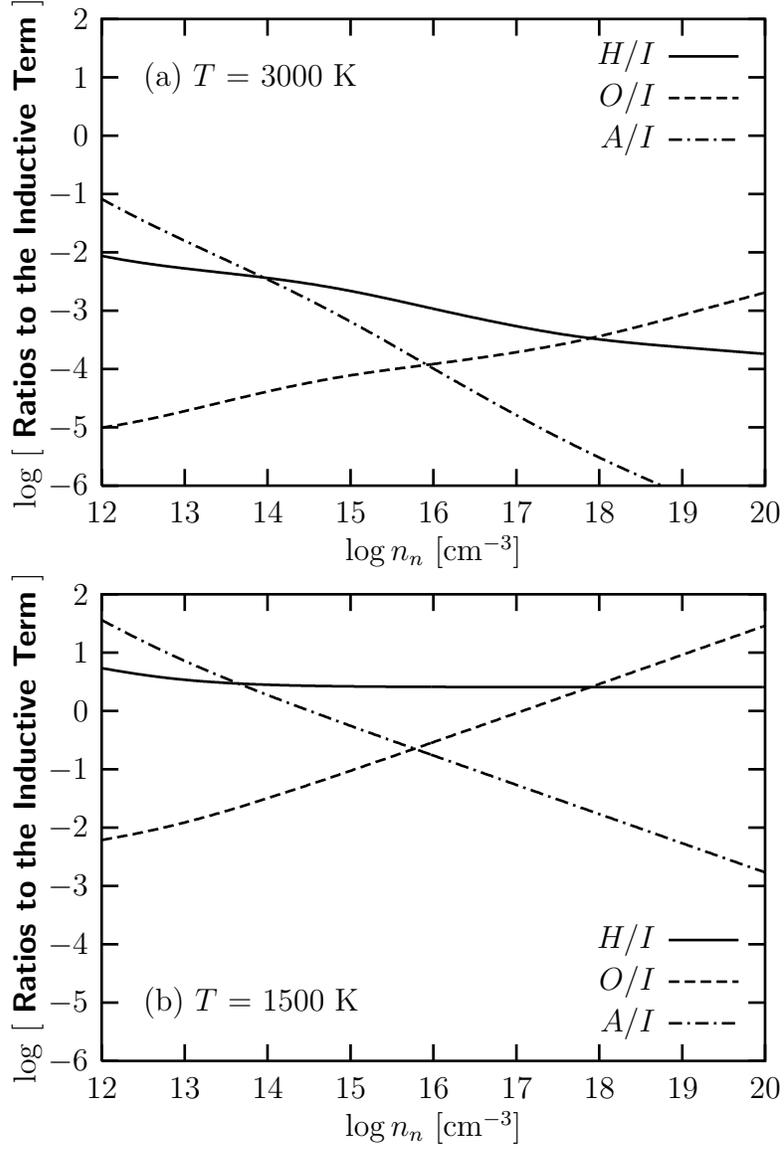

The ratios of non-ideal MHD terms to the inductive term in the
induction equation (\ref{eqn:faraday}) are functions of the ionization
fraction $n_e/n_n$.  The electron density $n_e$ in dwarf nova disks
comes from solving the Saha equation for Na, Al, and K assuming solar
abundances.
Figure \ref{fig:dn} shows the ratios $H/I$, $O/I$, and $A/I$ as a
function of the number density of neutrals $n_{n}$.  We assume $c_s /
v_{\rm A} = 1$ and $r = 10^{10}$ cm in this figure.  When the ratio
$H/I$ takes a value greater than unity, the Hall effect changes the
linear character of the MRI (Balbus \& Terquem 2001).  This ratio is
found to be very sensitive to the temperature of the disk.

For $T = 3000$ K, the typical value of $|X|$ in the Hall regime is
$10^{-3}$ at $n_n \sim 10^{18}$ cm$^{-3}$, so that the Hall term is
dominant, but
not large enough to affect the linear growth of the MRI.  
The magnetic Reynolds number $Re_{M} = (O/I)^{-1} \sim
10^4$ in this case, implying that ohmic dissipation is also unimportant
at this temperature.  
When $Re_{M} \lesssim 1$, ohmic dissipation reduces the linear growth
rate of the MRI (Sano \& Miyama 1999) and the amplitude of the Maxwell
stress at saturation (Sano \& Inutsuka 2001).  
Note that we use a different definition of $Re_{M}$ than Fleming et
al. (2000), which changes this critical value (see \S 5.2). 
The ratios shown in Figure \ref{fig:dn} depend on
the field strength, or $c_{s}/v_{\rm A}$ (see eqs. [\ref{eqn:ah2}] --
[\ref{eqn:oi2}]).  The Hall parameter increases
when $c_s / v_{\rm A}$ increases because $H / I = |X| / 2 \propto c_s /
v_{\rm A}$.  For the case with $c_s/v_{\rm A} = 1000$, the Hall
parameter is order unity, with a corresponding shift in
the density range of the Hall regime to $10^{10} \lesssim n_n \lesssim
10^{15}$ cm$^{-3}$.

For $T = 1500$ K, the Hall parameter $|X| \sim 1$, and the Hall term can
play an important role in the evolution of the MRI.  The magnetic
Reynolds number $Re_{M} = (O/I)^{-1} \sim 1$ -- 100 at this
temperature. 
Therefore, both the Hall effect and ohmic dissipation are
essential to the evolution of the MRI in dwarf nova disks when the
temperature is less than 1500 K.

\subsection{Protoplanetary Disks}

The temperatures of protoplanetary disks are very low, so that the
only sources of ionization are non-thermal, e.g., cosmic rays
(Umebayashi \& Nakano 1988; Gammie 1996), X-rays (Igea and Glassgold
1999), and the decay of radioactive elements.
We shall briefly investigate which non-ideal MHD effects are important
at different radii in the midplane of several disk models.

We assume a power-low distribution for the column density $\Sigma$ and
the temperature $T$,
\begin{equation}
\Sigma (r) = \Sigma_0 \left( \frac{r}{1 ~{\rm AU}} \right)^{-p_1}
\end{equation}
and
\begin{equation}
 T(r) = T_0 \left( \frac{r}{1 ~{\rm AU}} \right)^{-p_2} ~,
\end{equation}
where the mass of the central star is set to be $M = M_{\odot}$.  
The minimum-mass solar nebula is chosen as a fiducial model; $\Sigma_0
= 1.7 \times 10^3 ~{\rm g}~{\rm cm}^{-2}$, $p_1 = 3/2$, $T_0 = 280
~{\rm K}$, and $p_2 = 1/2$.  
The sound speed is $c_s(r) = [ k T(r) / \mu m_{\rm H}]^{1/2}$,
where $k$ is the Boltzmann constant and $\mu = 2.34$ is the mean
molecular weight.  The angular velocity is $\Omega(r)
= (G M_{\odot} / r^3)^{1/2}$, where $G$ is the gravitational constant.
For simplicity, cosmic rays and radioactive elements are considered as
the only sources of ionization.  The ionization rate due to cosmic
rays, $\zeta_{\rm CR}$, is dominant
because $\zeta_{CR}$ is about five orders of magnitude larger than the
ionization rate due to radioactive elements $\zeta_{\rm R}$.  But
cosmic rays cannot penetrate into the disk more than the attenuation
length $\chi_{\rm CR} \approx 96 ~{\rm g}~{\rm cm}^{-2}$ (Umebayashi \&
Nakano 1981).  Thus, the ionization rate at the midplane decreases
dramatically if the column density of the disk exceeds $\chi_{\rm
CR}$.  The ionization rate at the midplane is given by
\begin{equation}
 \zeta(r) \approx \zeta_{\rm CR} \exp \left[ -
\frac{\Sigma(r)}{2 \chi_{\rm CR}} \right]
+ \zeta_{\rm R} ~,
\label{eqn:zetaz}
\end{equation}
where $\zeta_{\rm CR} \approx 10^{-17} ~{\rm s}^{-1}$ and $\zeta_{\rm R}
\approx 6.9 \times 10^{-23} ~{\rm s}^{-1}$.

The ionization fraction at the ionization-recombination equilibrium is
approximately given by 
\begin{equation}
\frac{n_e}{n_n} = \sqrt{\frac{\zeta}{\beta n_n}} ~,
\end{equation}
where $\beta = 1.1 \times 10^{-7} (T/300 
~{\rm K})^{-1}$ is the dissociative recombination rate (Millar,
Farquhar, \& Willacy 1997).
The number density of electrons $n_e$ is a function of $n_n$ at the 
midplane, which is 
\begin{equation}
n_n(r) = \frac{\Sigma(r)}{\sqrt{\pi} \mu m_{\rm H} H(r)} ~,
\label{eqn:nn}
\end{equation}
where $H(r) = c_s(r) / \Omega(r)$ is the scale height of the disk.
For the magnetic field strength we simply assume $c_s / v_{\rm A} = 1$
or 10 at the midplane.

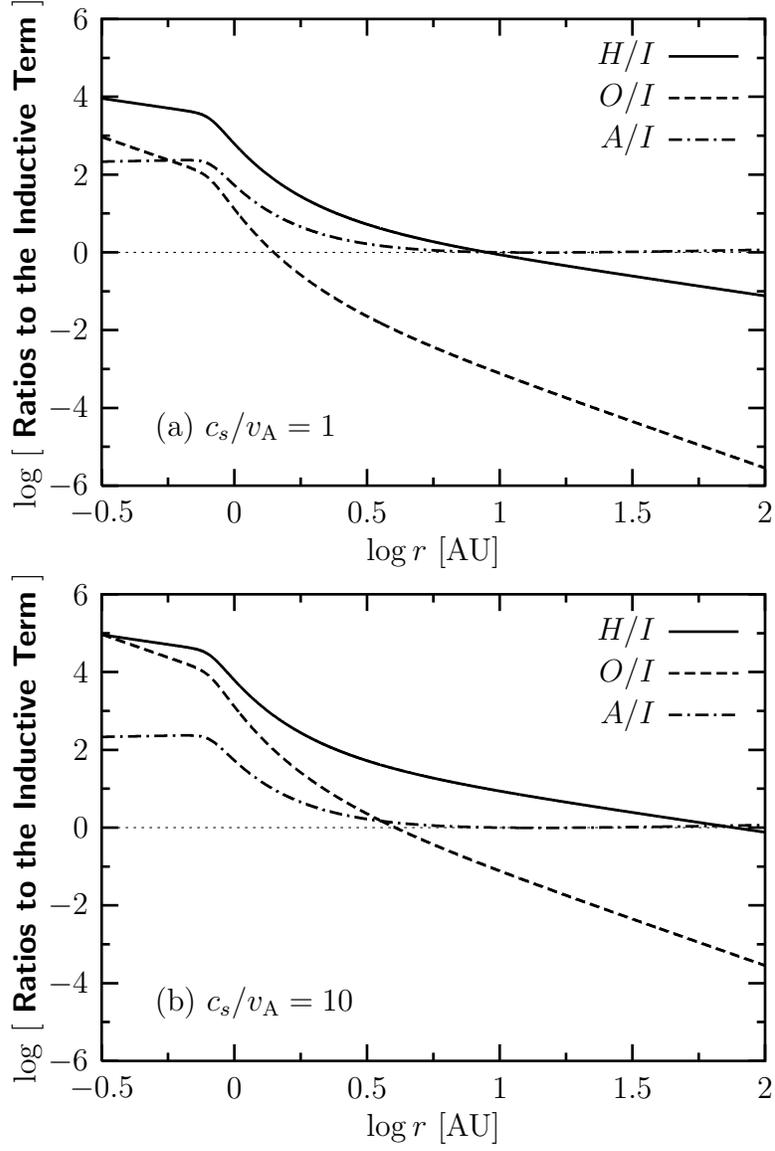
\begin{figure}
\begin{center}
\hspace*{-45pt}
\setlength{\unitlength}{0.1bp}%
{
}%
\begin{picture}(3600,2160)(0,0)%
{
}%
\put(3037,1605){\makebox(0,0)[r]{\large{${A}/{I}$}}}%
\put(3037,1762){\makebox(0,0)[r]{\large{${O}/{I}$}}}%
\put(3037,1919){\makebox(0,0)[r]{\large{${H}/{I}$}}}%
\put(1150,520){\makebox(0,0)[l]{\large (a) $c_s/v_{\rm A} = 1$}}%
\put(2200,50){\makebox(0,0){\large $\log r$ [AU]}}%
\put(700,1180){%
\makebox(0,0)[b]{\shortstack{\large{$\log$ [ {\bsf Ratios to the Inductive Term} ]}}}%
}%
\put(3450,200){\makebox(0,0)[c]{\large $2$}}%
\put(2950,200){\makebox(0,0)[c]{\large $1.5$}}%
\put(2450,200){\makebox(0,0)[c]{\large $1$}}%
\put(1950,200){\makebox(0,0)[c]{\large $0.5$}}%
\put(1450,200){\makebox(0,0)[c]{\large $0$}}%
\put(950,200){\makebox(0,0)[c]{\large $-0.5$}}%
\put(900,2060){\makebox(0,0)[r]{\large $6$}}%
\put(900,1767){\makebox(0,0)[r]{\large $4$}}%
\put(900,1473){\makebox(0,0)[r]{\large $2$}}%
\put(900,1180){\makebox(0,0)[r]{\large $0$}}%
\put(900,887){\makebox(0,0)[r]{\large $-2$}}%
\put(900,593){\makebox(0,0)[r]{\large $-4$}}%
\put(900,300){\makebox(0,0)[r]{\large $-6$}}%
\end{picture}%
\\ 
\hspace*{-45pt}
\setlength{\unitlength}{0.1bp}%
{
}%
\begin{picture}(3600,2160)(0,0)%
{
}%
\put(3037,1605){\makebox(0,0)[r]{\large{${A}/{I}$}}}%
\put(3037,1762){\makebox(0,0)[r]{\large{${O}/{I}$}}}%
\put(3037,1919){\makebox(0,0)[r]{\large{${H}/{I}$}}}%
\put(1150,520){\makebox(0,0)[l]{\large (b) $c_s/v_{\rm A} = 10$}}%
\put(2200,50){\makebox(0,0){\large $\log r$ [AU]}}%
\put(700,1180){%
\makebox(0,0)[b]{\shortstack{\large{$\log$ [ {\bsf Ratios to the Inductive Term} ]}}}%
}%
\put(3450,200){\makebox(0,0)[c]{\large $2$}}%
\put(2950,200){\makebox(0,0)[c]{\large $1.5$}}%
\put(2450,200){\makebox(0,0)[c]{\large $1$}}%
\put(1950,200){\makebox(0,0)[c]{\large $0.5$}}%
\put(1450,200){\makebox(0,0)[c]{\large $0$}}%
\put(950,200){\makebox(0,0)[c]{\large $-0.5$}}%
\put(900,2060){\makebox(0,0)[r]{\large $6$}}%
\put(900,1767){\makebox(0,0)[r]{\large $4$}}%
\put(900,1473){\makebox(0,0)[r]{\large $2$}}%
\put(900,1180){\makebox(0,0)[r]{\large $0$}}%
\put(900,887){\makebox(0,0)[r]{\large $-2$}}%
\put(900,593){\makebox(0,0)[r]{\large $-4$}}%
\put(900,300){\makebox(0,0)[r]{\large $-6$}}%
\end{picture}%
\\
\caption
{Ratios of non-ideal MHD terms to the inductive term given by equations
(\protect{\ref{eqn:rem}}) through (\protect{\ref{eqn:aoi}}) for the case
of a protoplanetary disk with (a) $c_s / v_{\rm A} = 1$ and (b) $c_s /
v_{\rm A} = 10$.
\label{fig:ppd}}
\end{center}
\end{figure}

Figure \ref{fig:ppd} shows the radial
distribution of the ratios $H/I$, $O/I$, and $A/I$ at the midplane for
the fiducial model with (a) $c_s/v_{\rm A} = 1$ and (b) $c_s/v_{\rm A}
= 10$.  When $c_s/v_{\rm A}=1$, the dominant term is the Hall effect
within 10 AU, while ambipolar diffusion dominates in the outer parts of
the disk.  When the collision frequency $\gamma \rho_i / \Omega =
(A/I)^{-1}$ is greater than unity, the disk is linearly unstable for
axisymmetric perturbations (Blaes \& Balbus 1994).  Thus the outer
region of the disk is marginally unstable to the MRI.

From equation (\ref{eqn:ho2}) the ratio $H/O$ is always larger than
unity when the density of neutrals is $n_n \lesssim 10^{18} (c_s /
v_{\rm A})^{-2}$ cm$^{-3}$.  
Note that this is independent of the ionization
fraction.  Thus the Hall effect is more important than ohmic
dissipation almost everywhere in protoplanetary disks.  The ratio $H/O$
becomes larger as the density decreases, or the distance from the star
$r$ increases.  The outer part of the disk has a higher ionization
fraction because of its lower density.  This means that both $H$ and
$O$ decrease as $r$ increases.  We define $r_O$ and $r_H$ as the radii
where the ratios $O/I$ and $H/I$ equal unity.  The Hall term is
important within $r < r_H$.

Protoplanetary disks can therefore be divided into three regions:  (1) $r > r_H$,
(2) $r_O < r < r_H$, and (3) $r < r_O$.  In the outer part of the disk
($r > r_H$), both the Hall effect and ohmic dissipation are unimportant
($|X| < 1$ and $Re_{M} \gg 1$).  The critical radius $r_H$ is about 10
AU in the fiducial model with $c_s / v_{\rm A} = 1$.  At intermediate
radii ($r_O < r < r_H$), the Hall term becomes important ($|X| > 1$)
but ohmic dissipation still can be ignored ($Re_M > 1$).  For the
region within $r_O$, the ionization fraction is so low that both
effects are important ($|X| \gg 1$ and $Re_{M} \ll 1$).  In the
fiducial model the critical radius $r_O \sim 1$ AU.  In the innermost
part of the disk $r \lesssim 0.1$ AU, thermal ionization is
efficient and the gas becomes well coupled to the magnetic field.

The critical radius $r_{H}$ is very sensitive to the field strength.
For example, almost all of the disk within 100 AU is in the Hall regime
for the case
$c_s / v_{\rm A} = 10$, as shown by Figure \ref{fig:ppd}b.  The Hall
parameter increases from 1 to $10^5$ as the radius $r$ decreases from
100 to 1 AU.  The change in the critical radius $r_{O}$ is small.
Within $r = r_O$ the magnetic Reynolds number is very small (1 --
$10^{-4}$).  The ratio $A/I$ is independent of $c_s / v_{\rm A}$.

The dependence of $r_{H}$ and $r_{O}$ on the distribution of the column
density is summarized in Table \ref{tbl:ppd}.  Disk models with $p_1 =
3 / 2$ and 1 are examined for two different disk masses $M_{\rm
disk}$.  The dependence of the critical radii on $p_1$ is very weak in
these cases.  As the disk mass is increased, the density at the midplane
increases, so that the ionization fraction decreases and
the critical radii shift outward in both cases.  The density at the
midplane also increases when the temperature decreases, because the
scale height of the disk $H$ is proportional to $T^{1/2}$ (see
eq. [\ref{eqn:nn}]).
Furthermore the recombination rate increases as the temperature
decreases.  Thus a lower temperature means a lower ionization
fraction.  If a constant temperature $T = 10$ K is assumed everywhere
in the disk, the critical radii become $r_O = 2.9$ AU and $r_H = 36$ AU
for the fiducial $\Sigma$ distribution with $c_s / v_{\rm A} = 1$.

In all of the models examined here, a large part of the disk ($r < r_{H} 
\sim 10$ -- 100 AU) is in the Hall regime, while the inner regions
($r < r_{O} \sim 1$ -- 5 AU) are very resistive ($Re_{M} < 1$).  
As discussed at the start of this section, this picture could be modified
by the effect of dust grains (Wardle \& Ng 1999; Sano et al. 2000).

\section{NUMERICAL METHOD}

\subsection{Equations and Algorithms}

We solve the equations of non-ideal MHD in a local Cartesian frame of
reference corotating with the disk at the angular frequency $\Omega$
corresponding to a fiducial radius $r_0$.  The coordinates in this
frame, written in terms of cylindrical coordinates ($r$, $\phi$, $z$),
are $x = r - r_0$, $y = r_0 (\phi - \Omega t)$, and $z$.  The vertical
component of gravity is ignored in this paper.  For a small region
$\Delta r$ surrounding the fiducial radius $r_0$, with $\Delta r \ll
r_0$, the basic equations are given by
\begin{equation}
\frac{\partial \rho}{\partial t} + 
\mbox{\boldmath $v$} \cdot \nabla \rho = 
- \rho \nabla \cdot \mbox{\boldmath $v$} ~,
\end{equation}
\begin{equation}
\frac{\partial \mbox{\boldmath $v$}}{\partial t} + 
\mbox{\boldmath $v$} \cdot \nabla \mbox{\boldmath $v$} = 
- \frac{\nabla P}{\rho} 
+ \frac{\mbox{\boldmath $J$} \times \mbox{\boldmath $B$}}{c \rho} 
- 2 \mbox{\boldmath $\Omega$} \times \mbox{\boldmath $v$} 
+ 2 q \Omega^2 x \hat{\mbox{\boldmath $x$}} ~,
\label{eqn:eom}
\end{equation}
\begin{equation}
\frac{\partial \epsilon}{\partial t} + 
\mbox{\boldmath $v$} \cdot \nabla \epsilon = 
- \frac{P \nabla \cdot \mbox{\boldmath $v$}}{\rho} +
\frac{4 \pi \eta \mbox{\boldmath $J$}^2}{c^2 \rho} ~,
\end{equation}
\begin{equation}
\frac{\partial \mbox{\boldmath $B$}}{\partial t} = 
\nabla \times \left( \mbox{\boldmath $v$} \times \mbox{\boldmath $B$} 
- \frac{4 \pi \eta \mbox{\boldmath $J$}}{c} -
\frac{\mbox{\boldmath $J$} \times \mbox{\boldmath $B$}}{e n_e}
\right) 
\equiv \nabla \times \mbox{\boldmath{$\cal{E}$}} ~,
\label{eqn:ineq}
\end{equation}
where $\epsilon$ is the specific internal energy and 
{\boldmath $\cal{E}$} is the electromotive force (EMF).  The term
$2 q \Omega^2 x$ in the equation of motion (\ref{eqn:eom}) is the tidal
expansion of the effective potential with a constant $q = 3 / 2$ for a
Keplerian disk.  The gas is assumed to be the ideal, with pressure $P =
( \gamma - 1 ) \rho \epsilon$ and $\gamma = 5 / 3$.  The induction equation
(\ref{eqn:ineq}) includes terms for both ohmic dissipation and the Hall
effect.

These equations are solved using a finite-difference code (Sano,
Inutsuka, \& Miyama 1999) in the local shearing box developed by
Hawley, Gammie, \& Balbus (1995).  The hydrodynamics module of our
scheme is based on a second order Godunov scheme (van Leer 1979) using
a nonlinear Riemann solver that is modified to account for the effect
of tangential magnetic fields.  The evolution of the magnetic field is
calculated with the Constrained Transport (CT) method (Evans \& Hawley
1988), which guarantees the divergence free condition of the field,
$\nabla \cdot \mbox{\boldmath $B$} = 0$, is maintained.  To extend the
method to include non-ideal MHD effects, we use an operator split
solution procedure.  The MoC-CT technique (Stone \& Norman 1992) is
used to update the inductive term $\mbox{\boldmath $v$} \times
\mbox{\boldmath $B$}$ in the EMF.  The algorithm to update the Hall
EMF, $ \mbox{\boldmath ${\cal E}$}_H = - \mbox{\boldmath $J$} \times
\mbox{\boldmath $B$}/e n_e$, is described in detail in the Appendix.

\subsection{Initial Conditions}

Our simulations begin with Keplerian shear flow, so that in the
corotating frame the azimuthal velocity is given by $v_{y0} = - q
\Omega x$.  Two initial field configuration are considered in this
paper: a uniform vertical field $B_z = B_0$, and a zero-net flux
vertical field $B_z(x) = B_0 \sin ( 2 \pi x / L_x)$, where $L_x$ is the
box size in the radial direction.  Except for the shear velocity, the
initial state is uniform: $\rho = \rho_0$ and $P = P_0$.

The calculations are performed in a two-dimensional region in the
radial-vertical ($x$-$z$) plane, in a volume bounded by $x = \pm H/2$ and
$z = \pm H/2$, where $H \equiv ( 2 / \gamma )^{1/2} c_{\rm s} /
\Omega$  is the scale height of the disk.  Most of the runs use a
standard grid resolution of $128 \times 128$ with uniform zoning.  In
the vertical direction, periodic boundary condition is used.  For the
radial boundary, a sheared periodic boundary condition (Hawley \&
Balbus 1992) is adopted.  We choose normalizations with $\rho_0 = 1$,
$H = 1$, and $\Omega = 10^{-3}$.  Then the sound velocity and gas
pressure are initially $2 c_{s0}^2 / \gamma = 10^{-6}$ and $P_0 = 5
\times 10^{-7}$, respectively.  Initial perturbations are introduced as
spatially uncorrelated pressure and velocity fluctuations.  These
fluctuations have a zero mean value with a maximum amplitude of $|
\delta P | / P_0 = 10^{-2}$ and $| \delta \mbox{\boldmath $v$} | / c_s
= 10^{-2}$.

This system is characterized by three non-dimensional parameters,
$\beta_0$, $Re_{M0}$, and $X_0$.  The initial plasma beta $\beta_0 =
(2/\gamma) c_{s0}^2 / v_{A0}^2$ measures the initial field strength,
where $v_{A0} = B_0/(4 \pi \rho_0)^{1/2}$.  The magnetic Reynolds number
$Re_{M0} = v_{A0}^2/\eta\Omega$ sets the value of the magnetic
diffusivity $\eta$, which is assumed to be spatially and temporally
constant in the calculation.  The initial Hall parameter,
\begin{equation}
 X_0 = \frac{cB_0\Omega}{2 \pi e n_{e0} v_{A0}^2} ~,
\end{equation}
indicates the importance of the Hall effect and determines the initial
number density of electrons $n_{e0}$.  We assume the electron abundance
is constant throughout the calculation and thus $n_e = n_{e0} \rho /
\rho_0$.  Our simulations show that the density fluctuations in the
nonlinear regime are not large, so that even if $n_e$ is assumed to be
constant (i.e., $n_e = n_{e0}$ throughout the simulation) there is
little difference to the results presented here.

\subsection{Linear Growth Rate of the Magnetorotational Instability}

To test the accuracy of our numerical algorithms, as well as to
summarize the effect of the Hall term on the linear dispersion relation
of the MRI, we first compare numerically measured growth rates with the
predictions of linear theory.  
Balbus \& Terquem (2001)
derive the dispersion relation for axisymmetric perturbations
$\delta \propto \exp (i k z + \sigma t)$ on
a uniform vertical field $B$.  Using the two non-dimensional
parameters $Re_M$ and $X$ defined by equations (\ref{eqn:rem}) and
(\ref{eqn:x}), the dispersion equation can be written as
\begin{equation}
\tilde{\sigma}^4 + \frac{2 \tilde{k}^2}{Re_{M}} \tilde{\sigma}^3
+ {\cal{C}}_2 \tilde{\sigma}^2 
+ \frac{2 \tilde{k}^2}{Re_{M}} 
\left( \tilde{\kappa}^2 + \tilde{k}^2 \right) \tilde{\sigma}
+ {\cal{C}}_0 = 0 ~,
\label{eqn:linear}
\end{equation}
\begin{equation}
{\cal{C}}_2 = 
\tilde{\kappa}^2 + 2 \tilde{k}^2 + \frac{\tilde{k}^4}{Re_M^2}
+ \frac{\tilde{k}^2}{4} X 
\left( \tilde{\kappa}^2 - 4 + \tilde{k}^2 X \right) ~,
\end{equation}
\begin{equation}
{\cal{C}}_0 =
\tilde{k}^2 \left( 1 + \frac{\tilde{\kappa}^2}{4} X \right)
\left[ \tilde{\kappa}^2 - 4 + \tilde{k}^2 \left( 1 + X \right) \right]
+ \frac{\tilde{\kappa}^2 \tilde{k}^4}{Re_M^2} ~,
\end{equation}
where $\tilde{\sigma} = \sigma / \Omega$, $\tilde{k} = k v_{\rm A} /
\Omega$, and $\tilde{\kappa} = \kappa / \Omega$ is the epicyclic
frequency.  A positive real root $\sigma$ implies an unstable
exponential growth of the mode.

Figure \ref{fig:dr}a shows the dispersion relations for $Re_{M} = 100$
and $X = 4$, 0, $-2$, and $-4$.  Since $Re_{M} \gg 1$, the effect on 
growth rates of the MRI
is dominated by the Hall term.  The sign of the Hall parameter $X$ is
determined by the direction of the magnetic field relative to
{\boldmath $\Omega$}.  Positive (negative) $X$ denotes that the field
direction is in the same (opposite) sense of {\boldmath $\Omega$}.
Modes with a smaller wavenumber than the critical value $k_{\rm crit}$
are unstable for the MRI.
When $X$ is positive the critical wavenumber decreases as
the Hall parameter increases.  The maximum growth rate is independent
of $X$ and has the value $\tilde{\sigma}_{\max} \approx 0.75$, the same
as the ideal MHD case.  For $X < 0$, the critical wavenumber
$\tilde{k}_{\rm crit} \rightarrow \infty$; the MRI has no
characteristic scale in this case.  On the other hand, for $X \le -4$,
no mode is linearly unstable to the axisymmetric MRI.  

The dispersion relations for $Re_{M} = 1$ and $X = 4$, 0, $-2$, and $-4$
are shown in Figure \ref{fig:dr}b.  Now the maximum growth rates are
significantly reduced by ohmic dissipation.  Fields
oriented such that $X$ is positive produce a larger growth rate than
those with negative $X$.  The critical wavenumber for negative $X$
remains finite in this case, and $X \le -4$ is always linearly stable
for all axisymmetric modes.  
Thus the larger
Hall parameter $|X|$ enhances the growth rate of the MRI in some case
($X \gg 1$ at least at small $k$) 
and completely suppresses any growth in other case ($X < -4$), 
depending on the field geometry.

Decreasing $Re_{M}$ further leads to
even smaller growth rates and critical wavenumbers.  For example, at
$Re_M = 0.1$ the maximum growth rate is $\tilde{\sigma}_{\max} = 0.11$
(0.037) for the case of $X = 2$ ($-2$), while the
critical wavenumber is $\tilde{k}_{\rm crit} = 0.21$ (0.12).  These
values are about an order of magnitude smaller than the ideal MHD case.

We have measured the growth rate of the MRI using numerical simulations
with uniform $B_z$, $\beta_0 = 400$, and 256 $\times$ 256 grid zones.
The simulations begin with very small random perturbations, $| \delta
P | / P_{\rm init} = 10^{-6}$ and $| \delta \mbox{\boldmath $v$} | /
c_{\rm s} = 10^{-6}$, so that any growing modes should be well
described by the linear analysis during the first a few orbits of
evolution.  A 2D Fourier decomposition of the radial velocity $v_x$ is
carried out at frequent time intervals, and the growth rates are
measured for modes with zero radial wavenumber, $k_x = 0$, between
orbits 2 and 2.5.  These numerical growth rates are shown in Figure
\ref{fig:dr} by open circles.  The
numerical growth rates reproduce the analytical dispersion relation
extremely well, each point is within a few percent of the predicted
linear value.  This provides a powerful validation of the numerical
method.

\section{NONLINEAR EVOLUTION OF THE MAGNETOROTATIONAL INSTABILITY}

\subsection{2D Simulations with Uniform Vertical Fields}

In the ideal MHD case, the 2D MRI on a uniform $B_z$ evolves into an
exponentially growing channel flow whose amplitude is unbounded (Hawley
\& Balbus 1992), although saturation at finite amplitude can occur when
ohmic dissipation is included (Sano et al. 1998).  We now examine how
these results are changed by inclusion of the Hall effect.

The initial field geometry for all models in this subsection is a
uniform vertical field $B_z = B_0$, where $B_0$ is constant.  Various
models with different $Re_{M0}$ and $X_0$ have been calculated and are
listed in Table \ref{tbl:z}.  The critical wavelength $\lambda_{\rm
crit} \equiv 2 \pi / k_{\rm crit}$ and the maximum growth rate
$\sigma_{\max}$ are obtained from the dispersion equation
(\ref{eqn:linear}).  Hereafter $\langle f \rangle$ denotes the
volume-averaged value of $f$, while $\langle \negthinspace \langle f
\rangle \negthinspace \rangle$ denotes the time- and volume-averaged
value.  The time averaging is taken over the last 20 orbits of the
calculation.

When the Hall parameter $X_0$ is positive, the nonlinear evolution of
the MRI is quite similar to the case $X_0 = 0$.  For all the models
with $Re_{M0} \gtrsim 1$ and $X_0 \ge 1$, the nonlinear evolution of
the MRI goes into the two-channel flow as in the ideal MHD cases.
Figure \ref{fig:z5} shows the time evolution of the magnetic field
lines in the $Re_{M0} = 10$ and $X_0 = 2$ run (model Z5).  The gray
color contours denote the radial velocity $v_x$ normalized by the
initial sound speed $c_{s0}$.  In this
figure, time is given in orbits $t_{\rm rot} = 2 \pi / \Omega$.  
In the linear phase (until about 2
orbits), the modes with the most unstable wavelength $\lambda_{\rm MRI}
\sim 2 \pi v_{\rm A} / \Omega$ dominate.  However, because many other
unstable modes are also growing, in the nonlinear regime they strongly
interact leading to a transient phase of MHD turbulence.

In the turbulent phase, the vertical component of the field is
amplified, and the characteristic length of the MRI in the vertical
direction becomes longer.  Thus, at late times the magnetic field and
velocity are dominated by larger structures.  The final state shown
by the last panel of Figure \ref{fig:z5} is quite similar to the ideal
MHD case (Hawley \& Balbus 1992).  The magnetic field lines are almost
horizontal, and two radial streams, one inward and the other outward,
occupy the computational box.  The increase of the magnetic energy
continues without saturation until it exceeds $10 P_0$ and the
calculation is stopped.

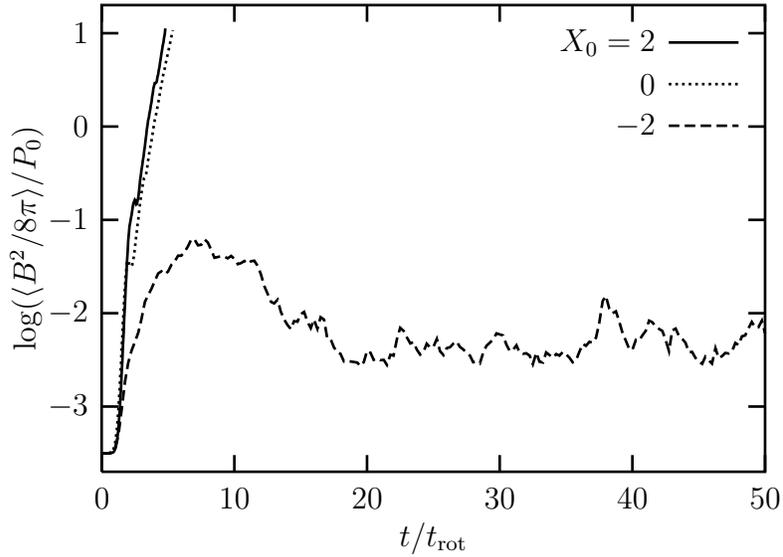
\begin{figure}
\begin{center}
\hspace*{-45pt}
\setlength{\unitlength}{0.1bp}%
{
}%
\begin{picture}(3600,2160)(0,0)%
{
}%
\put(3037,1605){\makebox(0,0)[r]{\large{$-2$}}}%
\put(3037,1762){\makebox(0,0)[r]{\large{$0$}}}%
\put(3037,1919){\makebox(0,0)[r]{\large{$X_0=2$}}}%
\put(2200,50){\makebox(0,0){\large $t / t_{\rm rot}$}}%
\put(700,1180){%
\makebox(0,0)[b]{\shortstack{\large{$\log (\langle B^2 / 8 \pi \rangle / P_0)$}}}%
}%
\put(3450,200){\makebox(0,0)[c]{\large $50$}}%
\put(2950,200){\makebox(0,0)[c]{\large $40$}}%
\put(2450,200){\makebox(0,0)[c]{\large $30$}}%
\put(1950,200){\makebox(0,0)[c]{\large $20$}}%
\put(1450,200){\makebox(0,0)[c]{\large $10$}}%
\put(950,200){\makebox(0,0)[c]{\large $0$}}%
\put(900,1954){\makebox(0,0)[r]{\large $1$}}%
\put(900,1602){\makebox(0,0)[r]{\large $0$}}%
\put(900,1250){\makebox(0,0)[r]{\large $-1$}}%
\put(900,898){\makebox(0,0)[r]{\large $-2$}}%
\put(900,546){\makebox(0,0)[r]{\large $-3$}}%
\end{picture}%
\\ 
\caption
{Time evolution of the volume-averaged magnetic energy $\langle
B^2 / 8 \pi \rangle / P_0$ for model Z5 ($X_0 = 2$), Z6 ($X_0 = 0$), and
Z7 ($X_0 = -2$). 
The plasma beta and the magnetic Reynolds number of these models are
$\beta_0 = 3200$ and $Re_{M0} = 10$.
\label{fig:em-t-rm10}}
\end{center}
\end{figure}

Figure \ref{fig:em-t-rm10} shows the time evolution of the
volume-averaged magnetic energy $\langle B^2 / 8 \pi \rangle$ for
models with $X_0 = 2$, 0, and $-2$ (model Z5, Z6, and Z7).  The
magnetic Reynolds number of these models is $Re_{M0} = 10$.  As
mentioned above, the $X_0 = 2$ and 0 runs show unbounded exponential
growth.  Interestingly, the $X_0 = -2$ run shows amplification of the
magnetic energy by an order of magnitude during the linear phase,
followed by saturation at finite amplitude.  This amplitude is
sustained for at least 50 orbits.

We find that the growth of the MRI saturates for all models with $X_0 <
0$.  Figure \ref{fig:z7} shows images of typical time evolution of
the field lines in a saturated model (Z7).  The angular momentum
perturbations $\delta v_y = v_y - v_{y0}$ are also shown by gray color
contours.  Obviously, smaller scale fluctuations are amplified in the
linear phase compared to the model with $X_0 = 2$ (shown in
Fig. \ref{fig:z5}).  Larger scale structures do not dominate at the
nonlinear stage for $X_0 < 0$, because small wavelength perturbations
are always unstable even after the field is amplified by the MRI (see \S
5.1).

The nonlinear stage of model Z7 is MHD turbulence as shown by the last
panel of Figure \ref{fig:z7}.  The efficiency of angular momentum
transport is measured by the turbulent stress $w_{xy}$, which is the
sum of the Maxwell stress, $- B_x B_y / 4 \pi$, and the Reynolds
stress, $\rho v_x \delta v_y$.  The $\alpha$ parameter of Shakura \&
Sunyaev (1973) can then be expressed as $\alpha = w_{xy} / P_0$.  
The Maxwell stress is well-correlated with the magnetic energy, so that 
the time evlution of the stress is similar to that of the magnetic
energy shown by Figure \ref{fig:em-t-rm10}.
We find that the saturation level of the Maxwell and Reynolds stresses
are almost the same in model Z7.  
The time- and volume-averaged stresses are
$\langle \negthinspace \langle - B_x B_y / 4 \pi \rangle \negthinspace
\rangle / P_0 = 1.4 \times 10^{-3}$ and $\langle \negthinspace \langle
\rho v_x \delta v_y \rangle \negthinspace \rangle / P_0 = 1.3 \times
10^{-3}$.  The turbulence in this case is slightly anisotropic:
the components of the magnetic and perturbed kinetic energy are
$\langle \negthinspace \langle B_y^2 \rangle
\negthinspace \rangle = 4.6 \langle \negthinspace \langle B_x^2 \rangle
\negthinspace \rangle = 1.8 \langle \negthinspace \langle B_z^2 \rangle
\negthinspace \rangle$ and $\langle \negthinspace \langle \rho \delta
v_y^2 \rangle \negthinspace \rangle = 1.2 \langle \negthinspace \langle
\rho v_x^2 \rangle \negthinspace \rangle = 1.4 \langle \negthinspace
\langle \rho v_z^2 \rangle \negthinspace \rangle$.

The effective Hall parameter in the nonlinear regime is defined as 
\begin{equation}
X_{\rm eff} \equiv \frac{c B_z \Omega}{2 \pi e n_e v_{\rm A}^2} 
= \frac{2 c \rho B_z \Omega}{e n_e B^2} ~.
\label{eqn:xeff}
\end{equation}
Because the Hall parameter is inversely proportional to the field
strength, the volume-averaged value $| \langle X_{\rm eff} \rangle |$
decreases as the field is amplified.  The time-averaged value is
$\langle \negthinspace \langle X_{\rm eff} \rangle \negthinspace
\rangle = - 0.16$ for model Z7, which started with $X_0 = -2$.  At the
end of model Z5, the volume-averaged effective Hall parameter is
$\langle X_{\rm eff} \rangle \sim 10^{-4}$.
The sign of $\langle X_{\rm eff} \rangle$ remains
unchanged for both models.  Thus the importance of the Hall term
becomes smaller as the MRI amplifies the field strength in the disk.
When the field is amplified, the magnetic Reynolds number at the
nonlinear stage $\langle v_{{\rm A}z}^2 / \eta \Omega \rangle$ 
increases, and then ohmic
dissipation also becomes less effective (Sano et al. 1998).

When $Re_{M0} < 1$, the growth of the MRI saturates even without the
Hall term (Sano et al. 1998).  Thus, we find the Hall term has less
effect in this case:  the nonlinear evolution with $X_0 = \pm 2$ is
quite similar to that of $X_0 = 0$.  Ohmic dissipation controls the MRI
in these cases, and the dependence on $X_0$ is very small except for
the linear growth rate.  Figure \ref{fig:by-t-rm0.1} shows the time
evolution of the azimuthal component of the magnetic energy $\langle
B_y^2 / 8 \pi \rangle$ for the models with $Re_{M0} = 0.1$ and $X_0 =
2$, 0, and $-2$ (model Z13, Z14, and Z15).  The linear growth phase
continues until a few tens of orbits because of the small growth rates
in these models.  The magnetic energy is amplified up to the initial
thermal energy $\sim P_0$ for all the models.  However the energy is
not sustained, and finally dies out after 100 orbits.  The Maxwell
stress at the end of the calculation is very small for these cases,
$\langle \negthinspace \langle - B_x B_y / 4 \pi \rangle \negthinspace
\rangle / P_0 \sim 10^{-9}$.

Although the magnetic field returns to its initial configuration after
100 orbits, we do not see the MRI re-emerge.  In 3D, re-emergence of
the MRI on the background field (which can never be dissipated) leads
to strong fluctuations in, e.g., the Maxwell stress (Fleming et
al. 2000).  In these 2D simulations, the magnetic and perturbed
kinetic energy is nearly comparable at the end of the linear growth
phase, so that the kinetic energy is $\langle \rho \delta v^2 / 2
\rangle \sim 0.2 P_0 \gg B_0^2 / 8 \pi$.  After that, the magnetic
energy starts decaying while the kinetic energy remains larger than its
initial value.  The kinetic energy is contained mostly in velocity
along the field (vertical) direction.  This flow is subthermal but
super Alfv{\'e}nic; this may suppress further growth of the MRI.

The evolution of the MRI on uniform vertical fields in 2D is summarized
by Figure \ref{fig:fate}.  The character of the nonlinear regime in the
2D parameter space defined by $Re_{M0}$ and $X_0$ is denoted by open
circles for simulations that result in unbounded growth of the channel
solution, filled circles for simulations that result in saturation and
eventual decay, and crosses for simulations that show no growth at
all.  We find that when $Re_{M0} \gtrsim 1$ and $X_0 \ge 0$, an
unbounded two-channel flow emerges in the nonlinear regime.  Saturation
at finite amplitude occurs when $X_0 < 0$ or $Re_{M0} \lesssim 1$.
Finally, if $X_0 < -4$, all modes are stable:  no linear growth can be
seen for the models with $X_0 = -5$ (model Z4, Z8, Z12, and Z16) shown
by crosses in the figure.

\subsection{2D Simulations with Zero-Net Flux Vertical Fields}

Next we consider models with a zero-net flux vertical field, $B_z(x) =
B_0 \sin (2 \pi x / L_x)$.  In the ideal MHD case with this initial
field, the MRI evolves into a transient phase of MHD turbulence that
eventually dies away (Hawley \& Balbus 1992).  For this initial field,
the Hall parameter is given by $X(x) = X_0/\sin (2 \pi x / L_x)$ where
$X_0 = 2 c \rho_0 \Omega / e n_{e0} B_0$.  Thus, the region $x < 0$ has
positive $X$ while the region $x > 0$ has negative $X$, and the minimum
of $|X(x)|$ is $X_0$.  Table
\ref{tbl:s} lists the models computed with a zero-net flux.  The initial 
plasma beta $\beta_0$, the critical wavelength $\lambda_{\rm
crit}$, and the maximum growth rate
$\sigma_{\max}$ are given for $B_z = B_0$ and $X = X_0$.

Figure \ref{fig:s2} shows images of the magnetic energy in model S2,
with $Re_{M0} = 100$ and $X_0 = 2$.  The velocity field is also shown
by arrows.  Most of the region with negative $X$ is stable because the
Hall parameter $X < -4$.  The growth of the MRI is seen mainly in the
region with positive $X$, as evident in the first panel of Figure
\ref{fig:s2}.  The growth in $x < 0$ gradually affects the
structure of the region $x > 0$.  The amplified magnetic field spreads
across the entire computational domain by 3 orbits, after that the
amplified magnetic energy is sustained for at least 100 orbits.

Figure \ref{fig:em-t-sin} shows the time evolution of the magnetic
energy for $Re_{M0} = 100$, 10, and 1 with and without the Hall term
($X_0 = 0$ and 2).  For the $Re_{M0} = 100$ and 10 runs, the magnetic
energy is sustained until 100  orbits.  In the turbulent state the
magnetic and the perturbed kinetic energy are equi-partitioned.  Since
the vertical component of the magnetic energy, which is the seed of the
MRI, continues decaying throughout the evolution, the activity of the
turbulence gradually weakens over time in both runs.  Therefore, active
MHD turbulence is a transient phenomenon in these cases.  When $Re_{M0}
= 100$, the Maxwell stress at the end of the calculation is $\langle
\negthinspace \langle - B_x B_y / 4 \pi \rangle \negthinspace \rangle /
P_0 = 1.6 \times 10^{-5}$ and $2.4 \times 10^{-8}$ in the $X_0 = 0$ and
2 run, respectively.  Only the toroidal field remains at late phases of
the evolution.  In 2D, a disk with a purely toroidal field is stable to
the MRI.  However, non-axisymmetric unstable modes could grow from the
toroidal field and sustain MHD turbulence in 3D.

The evolution of more resistive models with $Re_{M0} = 1$ and
$X_0 = 0$ and 2 is quite
different.  The magnetic energy is amplified for several orbits, but
then dies out.  The decay time of the magnetic energy is about 5
orbits.  The diffusion time $t_{\rm diff}$ of magnetic fields for a
length scale $l$ is given by
\begin{equation}
\frac{t_{\rm diff}}{t_{\rm rot}} = 
\frac{l^2}{\eta t_{\rm rot}} =
5.1 \times 10^2 Re_{M0} \left( \frac{\beta_0}{3200} \right)
\left( \frac{l}{H} \right)^2 ~.
\end{equation}
Thus the decay time corresponds to the diffusion time $t_{\rm diff}$
with $l \sim 0.1 H \sim \lambda_{\rm crit}$.

\begin{figure}
\begin{center}
\hspace*{-45pt}
\setlength{\unitlength}{0.1bp}%
{
}%
\begin{picture}(3600,2160)(0,0)%
{
}%
\put(3037,442){\makebox(0,0)[r]{\large{$\langle - B_x B_y / 4 \pi \rangle / P_0$}}}%
\put(3037,599){\makebox(0,0)[r]{\large{$\langle B_z^2 / 8 \pi \rangle / P_0$}}}%
\put(3037,756){\makebox(0,0)[r]{\large{$\langle B_y^2 / 8 \pi \rangle / P_0$}}}%
\put(3037,913){\makebox(0,0)[r]{\large{$\langle B_x^2 / 8 \pi \rangle / P_0$}}}%
\put(2225,50){\makebox(0,0){\large $t / t_{\rm rot}$}}%
\put(700,1180){%
\makebox(0,0)[b]{\shortstack{\large{$\log (\langle B^2 / 8 \pi \rangle / P_0)$ \& $\log (\langle - B_x B_y / 4 \pi \rangle / P_0)$}}}%
}%
\put(3450,200){\makebox(0,0)[c]{\large $10$}}%
\put(2960,200){\makebox(0,0)[c]{\large $8$}}%
\put(2470,200){\makebox(0,0)[c]{\large $6$}}%
\put(1980,200){\makebox(0,0)[c]{\large $4$}}%
\put(1490,200){\makebox(0,0)[c]{\large $2$}}%
\put(1000,200){\makebox(0,0)[c]{\large $0$}}%
\put(950,2060){\makebox(0,0)[r]{\large $0$}}%
\put(950,1708){\makebox(0,0)[r]{\large $-2$}}%
\put(950,1356){\makebox(0,0)[r]{\large $-4$}}%
\put(950,1004){\makebox(0,0)[r]{\large $-6$}}%
\put(950,652){\makebox(0,0)[r]{\large $-8$}}%
\put(950,300){\makebox(0,0)[r]{\large $-10$}}%
\end{picture}%
\\ 
\caption
{Time evolution of each component of the magnetic energy and the Maxwell
stress in model S8 ($\beta_0 = 3200$, $Re_{M0} = 1$, and $X_0 = 100$).
\label{fig:b-t-sin}}
\end{center}
\end{figure}
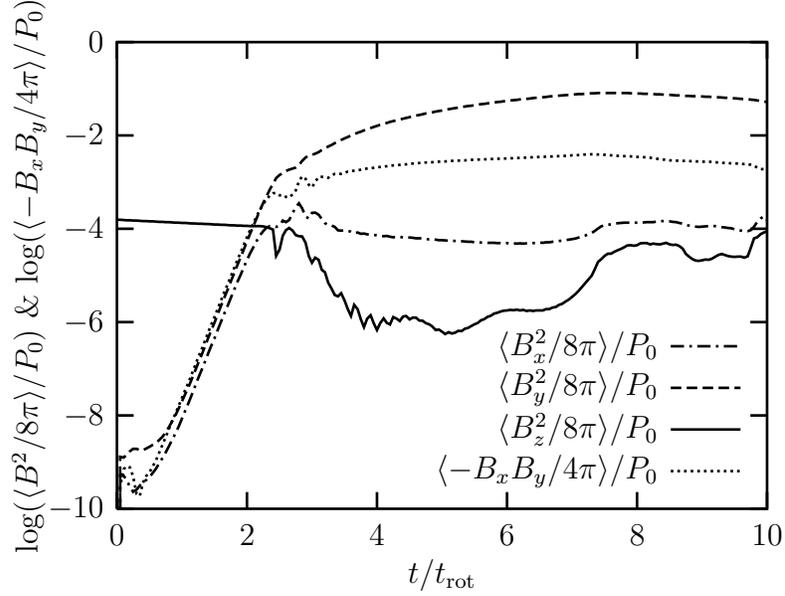

The Hall parameter can take values up to about $100/Re_{M0}$ (see
eq. [\ref{eqn:xrm}]) when the
disk is in the Hall regime.  Thus, a small magnetic Reynolds number
$Re_{M0}$ may give a very large Hall parameter $X_0$.  Since
the growth rate of the MRI is enhanced by large $X_0$ at small
$Re_{M0}$, it is important to examine whether significant growth can
occur in models with very large $X_0$.  Models S7 and S8 have $Re_{M0}
= 1$, and $X_0 = 10$ and 100 respectively.  The evolution of the $X_0 =
10$ run shows decay of the magnetic energy similar to the $X_0 = 2$ run
(model S6).  On the other hand, for $X_0 = 100$, active turbulence is
initiated and sustained for at least 10 orbits.  Figure
\ref{fig:b-t-sin} shows the time evolution of each component of the
magnetic field in this run.  
As in the other active turbulent models, the Maxwell stress is strongly
correlated with the magnetic energy and the time evolution of the stress
is resemble to that of the magnetic energy.
The stress is amplified up to $10^{-3} P_0$ and sustained until 10
orbits.
This saturation level is comparable to or slightly larger than those in
the less resistive models ($Re_{M0} = 100$, 10) at the same phase.

For more resistive models with $Re_{M0} = 0.1$, the nonlinear evolution in
all cases (models S9 -- S12) shows rapid decay of the magnetic field even
with very large
$X_0$.  The decay timescale of the magnetic energy is very short,
about an orbit.  In the $X_0 = 1000$ run with $Re_{M0} = 0.1$ (model
S12), the critical wavelength of the MRI is longer than the scale
height of the disk $H$, or the vertical box size.  No linear growth of
the MRI can be seen in the calculation, nor would the MRI be present
for these parameters in actual disks.

\section{DISCUSSION}

\subsection{Interpretation of the 2D Evolution}

\begin{figure}
\begin{center}
\hspace*{-45pt}
\setlength{\unitlength}{0.1bp}%
{
}%
\begin{picture}(3600,2160)(0,0)%
{
}%
\put(2225,50){\makebox(0,0){\large $X_0$}}%
\put(700,1180){%
\makebox(0,0)[b]{\shortstack{\large{$Re_{M0}$}}}%
}%
\put(3450,200){\makebox(0,0)[c]{\large $3$}}%
\put(3178,200){\makebox(0,0)[c]{\large $2$}}%
\put(2906,200){\makebox(0,0)[c]{\large $1$}}%
\put(2633,200){\makebox(0,0)[c]{\large $0$}}%
\put(2361,200){\makebox(0,0)[c]{\large $-1$}}%
\put(2089,200){\makebox(0,0)[c]{\large $-2$}}%
\put(1817,200){\makebox(0,0)[c]{\large $-3$}}%
\put(1544,200){\makebox(0,0)[c]{\large $-4$}}%
\put(1272,200){\makebox(0,0)[c]{\large $-5$}}%
\put(1000,200){\makebox(0,0)[c]{\large $-6$}}%
\put(950,1913){\makebox(0,0)[r]{\large $100$}}%
\put(950,1424){\makebox(0,0)[r]{\large $10$}}%
\put(950,936){\makebox(0,0)[r]{\large $1$}}%
\put(950,447){\makebox(0,0)[r]{\large $0.1$}}%
\end{picture}%
\\ 
\caption
{Final state for calculated disk models with a uniform vertical field.
Filled circles denote models in which the growth of the MRI
saturates as MHD turbulence in the nonlinear stage.
Open circles denote models which evolve into a channel flow that grows
without bound.
Crosses denote models which are linearly stable to the MRI.
The parameter space ($X_0$, $Re_{M0}$) is divided into four regions
shown by the dotted lines according to the characteristic of the linear
dispersion relation (see \S 5.1).
\label{fig:fate}}
\end{center}
\end{figure}
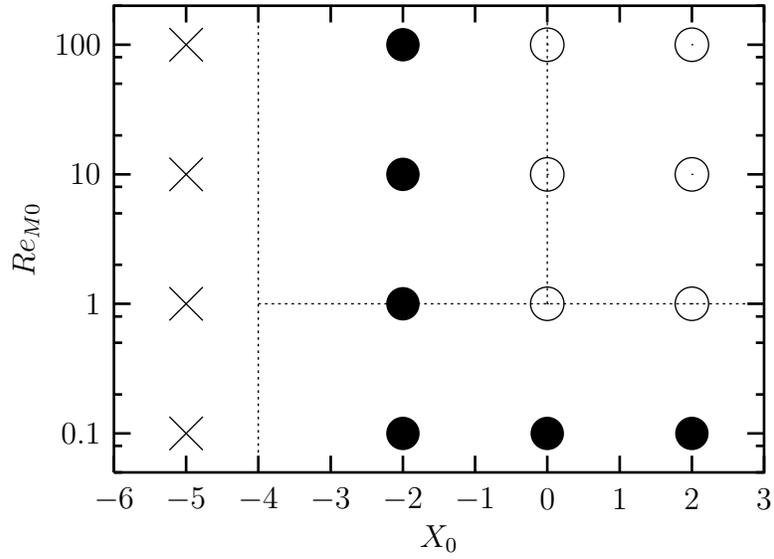

The axisymmetric MRI evolves into a channel flow or MHD turbulence,
depending on the parameters $Re_{M0}$ and $X_0$ (provided linearly
unstable modes exist at all).  The results summarized in Figure
\ref{fig:fate} reflect the linear properties of the MRI.  From the
dispersion relation, the parameter space $X_0 > -4$ can be divided into
three regions, as shown by dotted lines in Figure \ref{fig:fate}.  When
$Re_{M0} \gtrsim 1$ and $X_0 \ge 0$, the characteristic wavelength of
the MRI is proportional to $v_{A} / \Omega$, and a critical wavelength
exists.  For this case, the evolution of the MRI shows an inverse
cascade of the magnetic energy, and a two-channel flow emerges without
saturation.  This is because the characteristic scale increases as the
field strength is amplified by the growth of the MRI.  In the second
region ($Re_{M0} \lesssim 1$), the dispersion relation has a critical
wavelength, but the most unstable wavelength is proportional to $\eta /
v_{\rm A}$, that is inversely proportional to the field strength.  In
the third region ($X_0 < 0$), there is no characteristic scale of the
MRI.  Then smaller scale fluctuations are always unstable.  Therefore,
models in these last two regions show no evidence for an inverse
cascade; instead they evolve into MHD turbulence.

Models with $\beta = 800$ also show the same characteristics as the
models plotted in Figure \ref{fig:fate}, implying these results are
independent of the initial field strength.  Estimation of the saturated
amplitude of the Maxwell stress, which determines the efficiency of
angular momentum transport in actual accretion disks, requires 3D
simulations of the MRI.  In 3D simulations that include only the ohmic
dissipation, the saturation amplitude of the Maxwell stress is larger
if the MRI has an inverse cascade (Sano \& Inutsuka 2001).  Local 3D
simulations of the MRI in Hall MHD are presented in Sano \& Stone
(2002).

\subsection{Definition of Magnetic Reynolds Number}

In this paper we have defined the magnetic Reynolds number as $Re_{M} =
v_{\rm A}^2 / \eta \Omega$, and used $Re_{M0}$ to denote the value of
this number in the initial state.  Both a linear analysis of the MRI
including ohmic dissipation, as well as local 3D simulations 
with an initial uniform vertical field reveal
that the critical number required to generate significant MHD
turbulence and angular momentum transport is $Re_{M0} \sim 1$ (Sano \&
Inutsuka 2001).  If $Re_{M0} > 1$ the evolution and saturated state of
the MRI is little changed from the ideal MHD case, whereas if $Re_{M0}
< 1$ the growth rates and amplitude of the saturated state are both
significantly reduced compared to the ideal MHD case.

In actual accretion disks, there is a minimum value for $Re_{M}$ that
results from the requirement that the critical wavelength of unstable
modes in the very resistive regime $\lambda_{\rm MRI} \sim \eta /
v_{\rm A}$ be less than the disk thickness $H$ (Sano \& Miyama 1999). 
This requires $Re_{M0}
> v_{\rm A}/c_s$, which can also be used as a stability criterion
(Gammie 1996; Igea \& Glassgold 1999; Sano et al. 2000).  If the field
strength of accretion disks is subthermal and $v_{\rm A}/c_{s} < Re_{M0}
< 1$, the MRI will be present, but the growth rate, saturated amplitude,
and angular momentum transport rate will all be reduced compared to the
ideal MHD case.  In practice, the largest value for the ratio $v_{\rm
A}/c_s$ that can be reached in an accretion disk is probably unity
(otherwise the field will escape the disk via buoyancy).  In this case,
the magnetic Reynolds number becomes $Re_{M}' = c_s^2 / \eta \Omega$
(Gammie \& Menou 1998; Fleming et al. 2000), with the critical value
measured from 3D simulations about $10^{4}$.  This last form is
independent of the magnetic field strength in the disk, and therefore
can be measured more easily with observations.  However, even if the
observed $Re_{M}'$ is below the critical value $10^4$, the MRI may still
operate if $0.01 \lesssim v_{\rm A} / c_{s} < 1$.

\section{SUMMARY}

We have shown that in both dwarf nova disks in quiescence, and in
protoplanetary disks around young stars, the Hall term can be important
to the linear properties of the MRI, and therefore must be included in
realistic models of these systems.  Next, we investigated the
nonlinear evolution of the MRI in Hall MHD using axisymmetric 2D
simulations.  Although 3D simulations are essential for evaluating the
efficiency of angular momentum transport in accretion disks,
understanding the 2D evolution will be very helpful to the
interpretation of 3D results.  Fully 3D simulations of the nonlinear
regime of the MRI in Hall MHD are presented in a companion paper (Sano
\& Stone 2002).

Our findings are summarized as follows.
\begin{enumerate}
\item 
For dwarf nova disks in quiescence, both the Hall effect and ohmic
dissipation are very important when the temperature is less than about
1500 K.  The importance of these effects is very sensitive to the
temperature of the disk.

\item
In the region $r \lesssim r_{H}$ in protoplanetary disks, the Hall term
dominates over ohmic dissipation and ambipolar diffusion, assuming dust
grains have settled out to the midplane.
The critical radius $r_{H} \sim$ 10 -- 100
AU, and depends mainly on the magnetic field strength in the disk.
Inside a radius of $r_O \sim$ 1 -- 5 AU, ohmic dissipation suppresses
all modes of the MRI, except within about 0.1 AU where thermal
ionization becomes important.  If dust grains with the same size
distribution as interstellar grains are well mixed throughout the disk
(i.e., they have not settled to the midplane), $r_O$ increases
dramatically, and the entire disk is dominated by ohmic dissipation.

\item
For models with an initially uniform $B_z$, the MRI evolves into either
a two-channel flow which grows without bound (for $Re_{M0} \gtrsim 1$
{\em and} $X_0 \ge 0$), or saturates as MHD turbulence (for $Re_{M0} \lesssim
1$ {\em or} $X_0 < 0$).  If $Re_{M0} \lesssim 1$, saturation of the MRI
occurs, but the amplified magnetic field eventually dies out.

\item
For models with an initial zero-net flux vertical field, the MRI
saturates as MHD turbulence, which either is sustained or eventually
dies away depending on $Re_{M0}$.  In this case, the evolution is
determined mostly by the effect of ohmic dissipation and has little
dependence on the Hall parameter $X_0$.

\end{enumerate}

\acknowledgements
We thank Steven Balbus, Caroline Terquem, and Neal Turner for helpful
discussions.
Computations were carried out on VPP5000 at the National Astronomical
Observatory of Japan and VPP700 at the Subaru Telescope, NAOJ.

\appendix
\section{Algorithm for the Calculation of the Hall Term}

We adopt an operator split solution procedure for the update of the
magnetic field.  Following the constraint transport method (Evans \&
Hawley 1988), the Hall electromotive force (EMF) {\boldmath
$\cal{E}$}$_H$ is calculated at the zone boundaries.  For simplicity the
1D algorithm (in the $x$ direction) is described here; the extension
to multi-dimensions is straightforward.  In the 1D case, the tangential
components of the magnetic field and EMF are defined at the zone center
and zone boundary, respectively, and the $x$ component of the magnetic
field is constant $B_x$.

The change of the magnetic field due to the Hall EMF is given by
\begin{equation}
\frac{\partial \mbox{\boldmath $B$}}{\partial t} = \nabla \times
\mbox{\boldmath ${\cal{E}}$}_{H} = \nabla \times
\left[ -  Q_{H} \left( \nabla \times \mbox{\boldmath $B$} \right) 
\times \mbox{\boldmath $B$} \right] ~,
\label{eqn:hemf}
\end{equation}
where $Q_{H} = c / 4 \pi e n_{e}$ is assume to be constant.
Each component of the induction equation (\ref{eqn:hemf}) is written as
\begin{equation}
\frac{\partial B_y}{\partial t} = 
- \frac{\partial {\cal{E}}_{Hz}}{\partial x} =
Q_H B_x \frac{\partial^2 B_z}{\partial x^2} ~,
\label{eqn:by}
\end{equation}
\begin{equation}
\frac{\partial B_z}{\partial t} = 
\frac{\partial {\cal{E}}_{Hy}}{\partial x} =
- Q_H B_x \frac{\partial^2 B_y}{\partial x^2} ~,
\label{eqn:bz}
\end{equation}
where
\begin{equation}
{\cal E}_{Hy} = - Q_{H} B_x \frac{\partial B_y}{\partial x} ~,
\label{eqn:ey}
\end{equation}
\begin{equation}
{\cal E}_{Hz} = - Q_{H} B_x \frac{\partial B_z}{\partial x} ~.
\label{eqn:ez}
\end{equation}

For the update from $t^{n}$ to $t^{n+1}$, the partially
time-advanced values of the Hall EMF {\boldmath $\cal{E}$}$_H^{n+1/2}$
at $t^{n+1/2}$ are needed.
From equations (\ref{eqn:ey}) and (\ref{eqn:ez}), the time-advanced Hall 
EMF at a zone $x = i$ and $t^{n + 1/2}$ is given by
\begin{equation}
{\cal E}_{Hy,i}^{n+1/2} = - Q_{H} B_x 
\frac{B_{y,i}^{n+1/2} - B_{y,i-1}^{n+1/2}}{\Delta x} ~,
\label{eqn:byi}
\end{equation}
\begin{equation}
{\cal E}_{Hz,i}^{n+1/2} = - Q_{H} B_x 
\frac{B_{z,i}^{n+1/2} - B_{z,i-1}^{n+1/2}}{\Delta x} ~,
\label{eqn:bzi}
\end{equation}
where $\Delta x$ is the grid scale.
To evaluate $B^{n+1/2}$, we use equations (\ref{eqn:by}) and
(\ref{eqn:bz}); the induction equation including only the Hall
EMF.
Then the partial update of $B_{y}$ and $B_{z}$ is given by 
\begin{equation}
B_{y,i}^{n+1/2} = B_{y,i}^{n} + \frac{\Delta t} 2 Q_H B_{x} 
\frac{B_{z,i+1}^{n} - 2 B_{z,i}^{n} + B_{z,i-1}^{n}}{(\Delta x)^2} ~,
\end{equation}
\begin{equation}
B_{z,i}^{n+1/2} = B_{z,i}^{n} - \frac{\Delta t} 2 Q_H B_{x} 
\frac{B_{y,i+1}^{n} - 2 B_{y,i}^{n} + B_{y,i-1}^{n}}{(\Delta x)^2} ~,
\end{equation}
where $\Delta t$ is the time step.

Since our update of the Hall term is time explicit, we also must add a new
time step constraint.
The frequency of the whistler wave is approximately given by $\omega
\approx Q_H B_x k^2$, where $k$ is the wavenumber, thus the group
velocity is $2 Q_H B_x k$.
The new constraint on the time step is then
\begin{equation}
\Delta t \le \frac{\Delta x}{2 Q_H B_x k} \equiv 
\frac{(\Delta x)^2}{4 \pi Q_H B_x} ~,
\end{equation}
where we assume $k = 2 \pi / \Delta x$.

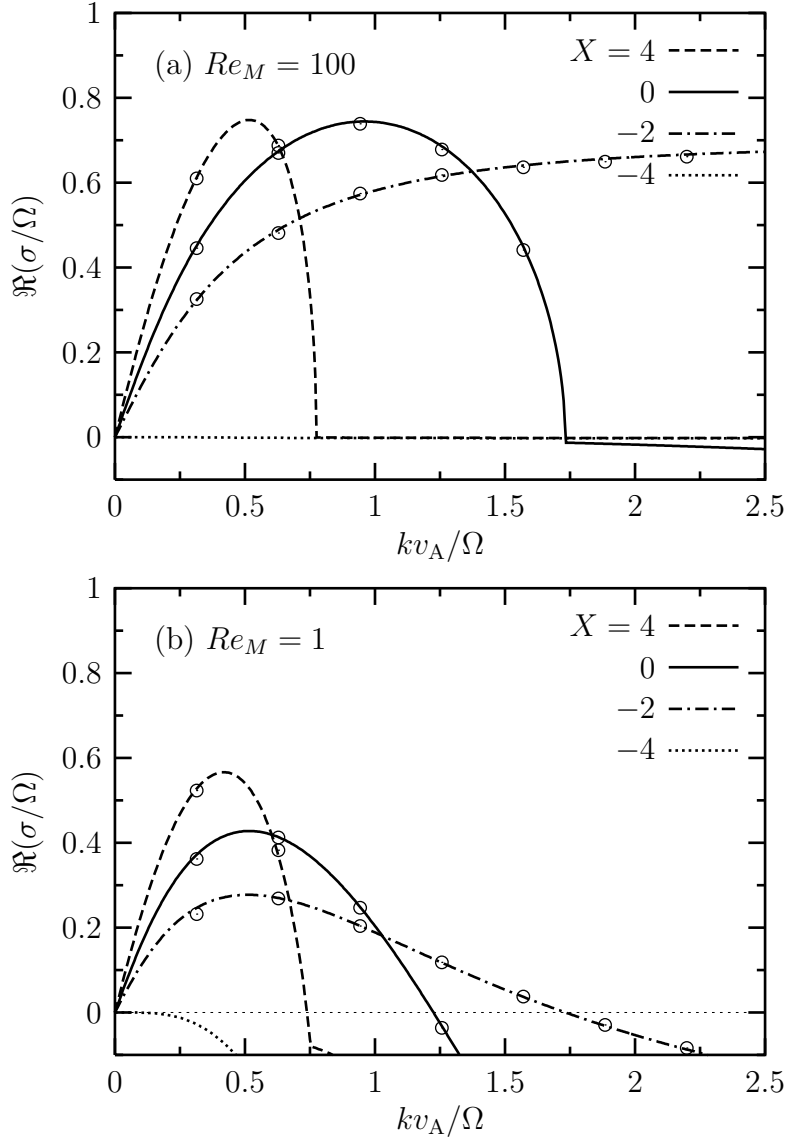
\begin{figure}
\begin{center}
\hspace*{-45pt}
\setlength{\unitlength}{0.1bp}%
{
}%
\begin{picture}(3600,2160)(0,0)%
{
}%
\put(3037,1448){\makebox(0,0)[r]{\large{$-4$}}}%
\put(3037,1605){\makebox(0,0)[r]{\large{$-2$}}}%
\put(3037,1762){\makebox(0,0)[r]{\large{$0$}}}%
\put(3037,1919){\makebox(0,0)[r]{\large{$X = 4$}}}%
\put(1147,1860){\makebox(0,0)[l]{\large (a) $Re_{M} = 100$}}%
\put(2225,50){\makebox(0,0){\large $k v_{\rm A} / \Omega$}}%
\put(700,1180){%
\makebox(0,0)[b]{\shortstack{\large{$\Re (\sigma / \Omega)$}}}%
}%
\put(3450,200){\makebox(0,0)[c]{\large $2.5$}}%
\put(2960,200){\makebox(0,0)[c]{\large $2$}}%
\put(2470,200){\makebox(0,0)[c]{\large $1.5$}}%
\put(1980,200){\makebox(0,0)[c]{\large $1$}}%
\put(1490,200){\makebox(0,0)[c]{\large $0.5$}}%
\put(1000,200){\makebox(0,0)[c]{\large $0$}}%
\put(950,2060){\makebox(0,0)[r]{\large $1$}}%
\put(950,1740){\makebox(0,0)[r]{\large $0.8$}}%
\put(950,1420){\makebox(0,0)[r]{\large $0.6$}}%
\put(950,1100){\makebox(0,0)[r]{\large $0.4$}}%
\put(950,780){\makebox(0,0)[r]{\large $0.2$}}%
\put(950,460){\makebox(0,0)[r]{\large $0$}}%
\end{picture}%
\\
\hspace*{-45pt}
\setlength{\unitlength}{0.1bp}%
{
}%
\begin{picture}(3600,2160)(0,0)%
{
}%
\put(3037,1448){\makebox(0,0)[r]{\large{$-4$}}}%
\put(3037,1605){\makebox(0,0)[r]{\large{$-2$}}}%
\put(3037,1762){\makebox(0,0)[r]{\large{$0$}}}%
\put(3037,1919){\makebox(0,0)[r]{\large{$X = 4$}}}%
\put(1147,1860){\makebox(0,0)[l]{\large (b) $Re_{M} = 1$}}%
\put(2225,50){\makebox(0,0){\large $k v_{\rm A} / \Omega$}}%
\put(700,1180){%
\makebox(0,0)[b]{\shortstack{\large{$\Re (\sigma / \Omega)$}}}%
}%
\put(3450,200){\makebox(0,0)[c]{\large $2.5$}}%
\put(2960,200){\makebox(0,0)[c]{\large $2$}}%
\put(2470,200){\makebox(0,0)[c]{\large $1.5$}}%
\put(1980,200){\makebox(0,0)[c]{\large $1$}}%
\put(1490,200){\makebox(0,0)[c]{\large $0.5$}}%
\put(1000,200){\makebox(0,0)[c]{\large $0$}}%
\put(950,2060){\makebox(0,0)[r]{\large $1$}}%
\put(950,1740){\makebox(0,0)[r]{\large $0.8$}}%
\put(950,1420){\makebox(0,0)[r]{\large $0.6$}}%
\put(950,1100){\makebox(0,0)[r]{\large $0.4$}}%
\put(950,780){\makebox(0,0)[r]{\large $0.2$}}%
\put(950,460){\makebox(0,0)[r]{\large $0$}}%
\end{picture}%
\\
\caption
{Dispersion relation for axisymmetric MRI modes for the case with (a)
$Re_{M} = 100$ and $Re_{M} = 1$.
Open circles depict the numerical growth rates measured in 2D numerical
simulations.
\label{fig:dr}}
\end{center}
\end{figure}

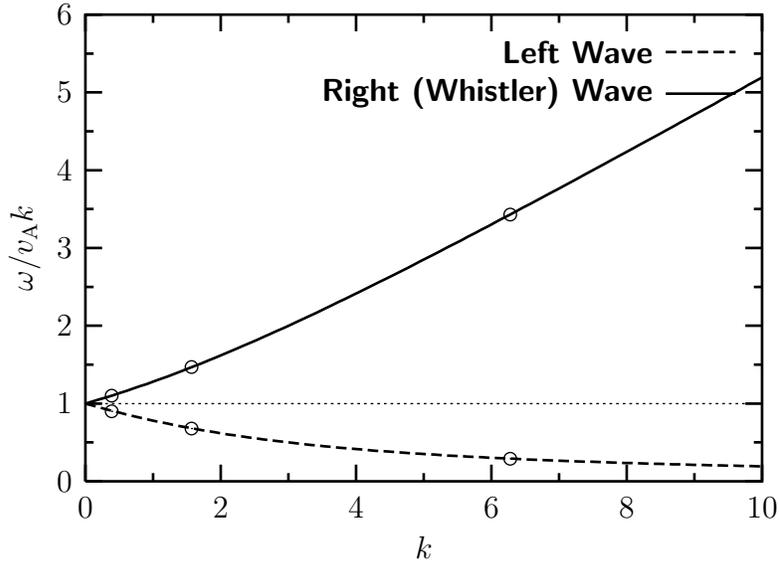
\begin{figure}
\begin{center}
\hspace*{-45pt}
\setlength{\unitlength}{0.1bp}%
{
}%
\begin{picture}(3600,2160)(0,0)%
{
}%
\put(3037,1762){\makebox(0,0)[r]{\large{\bsf Right (Whistler) Wave}}}%
\put(3037,1919){\makebox(0,0)[r]{\large{\bsf Left Wave}}}%
\put(2175,50){\makebox(0,0){\large $k$}}%
\put(700,1180){%
\makebox(0,0)[b]{\shortstack{\large{${\omega}/{v_{\rm A} k}$}}}%
}%
\put(3450,200){\makebox(0,0)[c]{\large $10$}}%
\put(2940,200){\makebox(0,0)[c]{\large $8$}}%
\put(2430,200){\makebox(0,0)[c]{\large $6$}}%
\put(1920,200){\makebox(0,0)[c]{\large $4$}}%
\put(1410,200){\makebox(0,0)[c]{\large $2$}}%
\put(900,200){\makebox(0,0)[c]{\large $0$}}%
\put(850,2060){\makebox(0,0)[r]{\large $6$}}%
\put(850,1767){\makebox(0,0)[r]{\large $5$}}%
\put(850,1473){\makebox(0,0)[r]{\large $4$}}%
\put(850,1180){\makebox(0,0)[r]{\large $3$}}%
\put(850,887){\makebox(0,0)[r]{\large $2$}}%
\put(850,593){\makebox(0,0)[r]{\large $1$}}%
\put(850,300){\makebox(0,0)[r]{\large $0$}}%
\end{picture}%
\\
\caption
{Dispersion relation of the left- and right-circularly polarized wave
traveling along field lines.
The Alfv{\'e}n wave for the case without the Hall effect ($\omega /
v_{\rm A} k = 1$) is shown by dotted line for comparison. 
Open circles denote numerically measured phase velocity in 1D
simulations.
\label{fig:lr}}
\end{center}
\end{figure}

In addition to checking the numerical growth rates of the MRI in Hall MHD 
with the predictions of linear theory (Fig. \ref{fig:dr}), we have
also tested our algorithm by comparing the numerically measured phase
velocity of whistler waves with the linear dispersion relation.
For this test, we adopt a uniform density $\rho$ and a uniform magnetic field
parallel to the $x$ axis, $B_x$ ($B_y = B_z = 0$), and consider the
propagation of linear waves proportional to $\exp i (k x -
\omega t)$ in the incompressible limit.
The dispersion equation without the Hall effect is $\omega^2 =
v_{\rm A}^2 k^2$ where $v_{\rm A}^2 = B_x^2 / 4 \pi \rho$.
If the Hall term is included, the dispersion equation is then written as 
\begin{equation}
( \omega ^2 - v_{\rm A}^2 k^2 )^2 = Q_{H}^2 B_x^2 k^4 \omega^2 ~.
\label{eqn:dr}
\end{equation}
Figure \ref{fig:lr} shows the dispersion relation obtained from
equation (\ref{eqn:dr}).  Due to the Hall effect, the Alfv{\'e}n wave
($\omega / v_{\rm A} k = 1$) splits into left- and right-circularly
polarized waves.  The right (whistler) wave can propagate much faster
than the Alfv{\'e}n wave.  Numerically obtained phase velocities are
also shown in the figure by open circles.  Each mode is calculated
using a 1D simulation of the propagation of a linear wave with 32 grid
zones per a wavelength.  The numerically measured phased velocities
agree with the dispersion relation very well (better than 1\%).  This
shows that our algorithm accurately captures the characteristics of
Hall MHD.

\clearpage 

\clearpage 

\begin{deluxetable}{ccccccc}
\tablecaption{Critical Radii in Models of Protoplanetary
 Disk\tablenotemark{a}}
\tablehead{
\colhead{} & \colhead{} & \colhead{} &
\multicolumn{2}{c}{$c_s/v_{A} = 1$} &
\multicolumn{2}{c}{$c_s/v_{A} = 10$} \\
\cline{4-5} \cline{6-7} 
\colhead{$\Sigma_0$} & \colhead{$p_1$} &
\colhead{$M_{\rm disk}$ [$M_{\odot}$]\tablenotemark{b}} &
\colhead{$r_{O}$ [AU]} & \colhead{$r_{H}$ [AU]} &
\colhead{$r_{O}$ [AU]} & \colhead{$r_{H}$ [AU]}
}
\startdata
$1.7 \times 10^3$ & $3/2$ & 0.024 & 1.4 & 8.9 & 4.0 & 76 \\
$1.7 \times 10^4$ & $3/2$ & 0.24  & 4.7 & 16  & 9.7 & 80 \\
$3.3 \times 10^2$ & $1$   & 0.024 & 0.58 & 8.4 & 2.9 & 76 \\
$3.3 \times 10^3$ & $1$   & 0.24  & 2.5 & 14 & 7.4 & 84 \\
\enddata
\tablenotetext{a}{The temperature distribution is assumed to be $T_0 = 280$ K
 and $p_2 = 1/2$ for all models.}
\tablenotetext{b}{The disk mass $M_{\rm disk}$ is integrated from 0.1 to
 100 AU.}
\label{tbl:ppd}
\end{deluxetable}

\begin{deluxetable}{ccccccccc}
\tablecaption{Uniform $B_z$ Simulations}
\tablehead{
\colhead{Model} & \colhead{Size} & 
\colhead{Grid} & \colhead{$\beta_0$} & 
\colhead{$Re_{M0}$} & \colhead{$X_0$} & 
\colhead{$\lambda_{\rm crit}/H$} & \colhead{$\sigma_{\max}/\Omega$} & 
\colhead{Orbits}
}
\startdata
Z1 & $H \times H$ & $128 \times 128$ & 3200 & 100 & 2  & 0.11 & 0.75 & 4.5 \\
Z2 & $H \times H$ & $128 \times 128$ & 3200 & 100 & 0  & 0.064 & 0.74 & 5.9 \\
Z3 & $H \times H$ & $128 \times 128$ & 3200 & 100 & $-2$ & 0.0 & 0.70 & 50 \\
Z4 & $H \times H$ & $128 \times 128$ & 3200 & 100 & $-5$ & \nodata &
 \nodata & 50 \\
Z5 & $H \times H$ & $128 \times 128$ & 3200 & 10 & 2  & 0.11 & 0.72 & 5.0 \\
Z6 & $H \times H$ & $128 \times 128$ & 3200 & 10 & 0  & 0.064 & 0.70 & 5.4 \\
Z7 & $H \times H$ & $128 \times 128$ & 3200 & 10 & $-2$ & 0.0 & 0.60 & 50 \\
Z8 & $H \times H$ & $128 \times 128$ & 3200 & 10 & $-5$ & \nodata &
 \nodata & 50 \\
Z9 & $H \times H$ & $128 \times 128$ & 3200 & 1 & 2  & 0.12 & 0.51 & 6.4
 \\
Z10 & $H \times H$ & $128 \times 128$ & 3200 & 1 & 0  & 0.091 & 0.43 &
 9.7 \\
Z11 & $H \times H$ & $128 \times 128$ & 3200 & 1 & $-2$ & 0.064 & 0.28 & 50 \\
Z12 & $H \times H$ & $128 \times 128$ & 3200 & 1 & $-5$ & \nodata &
 \nodata & 50 \\
Z13 & $2H \times 2H$ & $128 \times 128$ & 3200 & 0.1 & 2  & 0.54 & 0.11
 & 100 \\
Z14 & $2H \times 2H$ & $128 \times 128$ & 3200 & 0.1 & 0  & 0.64 & 0.074
 & 100 \\
Z15 & $2H \times 2H$ & $128 \times 128$ & 3200 & 0.1 & $-2$ & 0.91 &
 0.037 & 100 \\
Z16 & $2H \times 2H$ & $128 \times 128$ & 3200 & 0.1 & $-5$ & \nodata &
 \nodata & 100 \\
\enddata
\label{tbl:z}
\end{deluxetable}

\begin{deluxetable}{ccccccccc}
\tablecaption{Zero-Net Flux $B_z$ Simulations}
\tablehead{
\colhead{Model} & \colhead{Size} & 
\colhead{Grid} & \colhead{$\beta_0$} & 
\colhead{$Re_{M0}$} & \colhead{$X_0$} & 
\colhead{$\lambda_{\rm crit}/H$} & \colhead{$\sigma_{\max}/\Omega$} & 
\colhead{Orbits}
}
\startdata
S1 & $H \times H$ & $128 \times 128$ & 3200 & 100 & 0  & 0.064 & 0.74 &
 100 \\
S2 & $H \times H$ & $128 \times 128$ & 3200 & 100 & 2  & 0.11 & 0.75 &
 100 \\
S3 & $H \times H$ & $128 \times 128$ & 3200 & 10 & 0  & 0.064 & 0.70 &
 100 \\
S4 & $H \times H$ & $128 \times 128$ & 3200 & 10 & 2  & 0.11 & 0.72 &
 100 \\
S5 & $H \times H$ & $128 \times 128$ & 3200 & 1 & 0  & 0.091 & 0.43 &
 100 \\
S6 & $H \times H$ & $128 \times 128$ & 3200 & 1 & 2  & 0.12 & 0.51 & 100
 \\
S7 & $H \times H$ & $128 \times 128$ & 3200 & 1 & 10  & 0.22 & 0.64 & 100
 \\
S8 & $H \times H$ & $128 \times 128$ & 3200 & 1 & 100  & 0.65 & 0.74 & 10
 \\
S9 & $H \times H$ & $128 \times 128$ & 3200 & 0.1 & 0  & 0.64 & 0.074
 & 50 \\
S10 & $H \times H$ & $128 \times 128$ & 3200 & 0.1 & 10  & 0.40 & 0.23
 & 50 \\
S11 & $H \times H$ & $128 \times 128$ & 3200 & 0.1 & 100  & 0.66 & 0.62
 & 50 \\
S12 & $H \times H$ & $128 \times 128$ & 3200 & 0.1 & 1000  & 2.0 & 0.74
 & 50 \\
\enddata
\label{tbl:s}
\end{deluxetable}

\clearpage

\clearpage

\begin{figure}
\caption
{Magnetic field lines ({\it solid}) and radial velocity, $\delta v_x
/ c_{s0}$ ({\it gray color}) in model Z5 ($\beta_0 = 3200$, $Re_{M0} =
10$, and $X_0 = 2$) at orbits 1.5, 2, 3, and 5. 
\label{fig:z5}}
\end{figure}

\clearpage

\begin{figure}
\caption
{Magnetic field lines ({\it solid}) and angular momentum perturbation,
$\delta v_y / c_{s0}$ ({\it gray color}) in model Z5 ($\beta_0 = 3200$,
$Re_{M0} = 10$, and $X_0 = -2$) at orbits 2, 3, 5 and 10.
\label{fig:z7}}
\end{figure}

\begin{figure}
\begin{center}
\hspace*{-45pt}
\setlength{\unitlength}{0.1bp}%
{
}%
\begin{picture}(3600,2160)(0,0)%
{
}%
\put(3037,1605){\makebox(0,0)[r]{\large{$-2$}}}%
\put(3037,1762){\makebox(0,0)[r]{\large{$0$}}}%
\put(3037,1919){\makebox(0,0)[r]{\large{$X_0=2$}}}%
\put(2225,50){\makebox(0,0){\large $t / t_{\rm rot}$}}%
\put(700,1180){%
\makebox(0,0)[b]{\shortstack{\large{$\log (\langle B_y^2 / 8 \pi \rangle / P_0)$}}}%
}%
\put(3450,200){\makebox(0,0)[c]{\large $100$}}%
\put(2960,200){\makebox(0,0)[c]{\large $80$}}%
\put(2470,200){\makebox(0,0)[c]{\large $60$}}%
\put(1980,200){\makebox(0,0)[c]{\large $40$}}%
\put(1490,200){\makebox(0,0)[c]{\large $20$}}%
\put(1000,200){\makebox(0,0)[c]{\large $0$}}%
\put(950,1913){\makebox(0,0)[r]{\large $0$}}%
\put(950,1620){\makebox(0,0)[r]{\large $-2$}}%
\put(950,1327){\makebox(0,0)[r]{\large $-4$}}%
\put(950,1033){\makebox(0,0)[r]{\large $-6$}}%
\put(950,740){\makebox(0,0)[r]{\large $-8$}}%
\put(950,447){\makebox(0,0)[r]{\large $-10$}}%
\end{picture}%
\\ 
\caption
{Time evolution of the azimuthal component of the magnetic energy 
$\langle B_y^2 / 8 \pi \rangle / P_0$ for model Z13 ($X_0 = 2$), Z14
($X_0 = 0$), and Z15 ($X_0 = -2$).
The plasma beta and the magnetic Reynolds number of these models are
$\beta_0 = 3200$ and $Re_{M0} = 0.1$.
\label{fig:by-t-rm0.1}}
\end{center}
\end{figure}
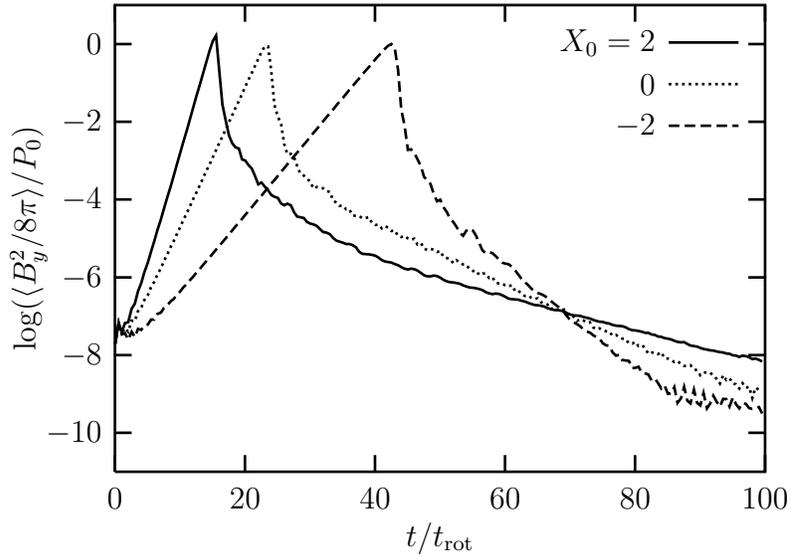

\clearpage

\begin{figure}
\caption
{Magnetic field energy, $\log (B^2 / 8 \pi P_0)$ ({\it gray color}) and
velocity field ({\it arrows}) in model S2 ($\beta_0 = 3200$, $Re_{M0} =
100$, and $X_0 = 2$) at orbits 1.5, 2, 3, and 5.
\label{fig:s2}}
\end{figure}

\begin{figure}
\begin{center}
\hspace*{-45pt}
\setlength{\unitlength}{0.1bp}%
{
}%
\begin{picture}(3600,2160)(0,0)%
{
}%
\put(3037,1762){\makebox(0,0)[r]{\large{$0$}}}%
\put(3037,1919){\makebox(0,0)[r]{\large{$X_0=2$}}}%
\put(1858,1792){\makebox(0,0)[c]{\large $Re_{M0}=10,100$}}%
\put(2495,560){\makebox(0,0)[r]{\large $Re_{M0}=1$}}%
\put(2960,1074){\makebox(0,0)[c]{\large $Re_{M0}=10$}}%
\put(2225,1532){\makebox(0,0)[l]{\large $Re_{M0}=100$}}%
\put(2225,50){\makebox(0,0){\large $t / t_{\rm rot}$}}%
\put(700,1180){%
\makebox(0,0)[b]{\shortstack{\large{$\log (\langle B^2 / 8 \pi \rangle / P_0)$}}}%
}%
\put(3450,200){\makebox(0,0)[c]{\large $100$}}%
\put(2960,200){\makebox(0,0)[c]{\large $80$}}%
\put(2470,200){\makebox(0,0)[c]{\large $60$}}%
\put(1980,200){\makebox(0,0)[c]{\large $40$}}%
\put(1490,200){\makebox(0,0)[c]{\large $20$}}%
\put(1000,200){\makebox(0,0)[c]{\large $0$}}%
\put(950,2060){\makebox(0,0)[r]{\large $0$}}%
\put(950,1708){\makebox(0,0)[r]{\large $-2$}}%
\put(950,1356){\makebox(0,0)[r]{\large $-4$}}%
\put(950,1004){\makebox(0,0)[r]{\large $-6$}}%
\put(950,652){\makebox(0,0)[r]{\large $-8$}}%
\put(950,300){\makebox(0,0)[r]{\large $-10$}}%
\end{picture}%
\\
\caption
{Time evolution of the magnetic energy $\langle \negthinspace \langle
B^2 / 8 \pi \rangle \negthinspace \rangle / P_0$ for model S1 ($Re_{M0}
= 100$ and $X_0 = 0$), S2 ($Re_{M0} = 100$ and $X_0 = 2$), S3 ($Re_{M0}
= 10$ and $X_0 = 0$), S4 ($Re_{M0} = 10$ and $X_0 = 2$), S5 ($Re_{M0}
= 1$ and $X_0 = 0$), and S6 ($Re_{M0} = 1$ and $X_0 = 2$).
The plasma beta of these models is $\beta_0 = 3200$.
\label{fig:em-t-sin}}
\end{center}
\end{figure}
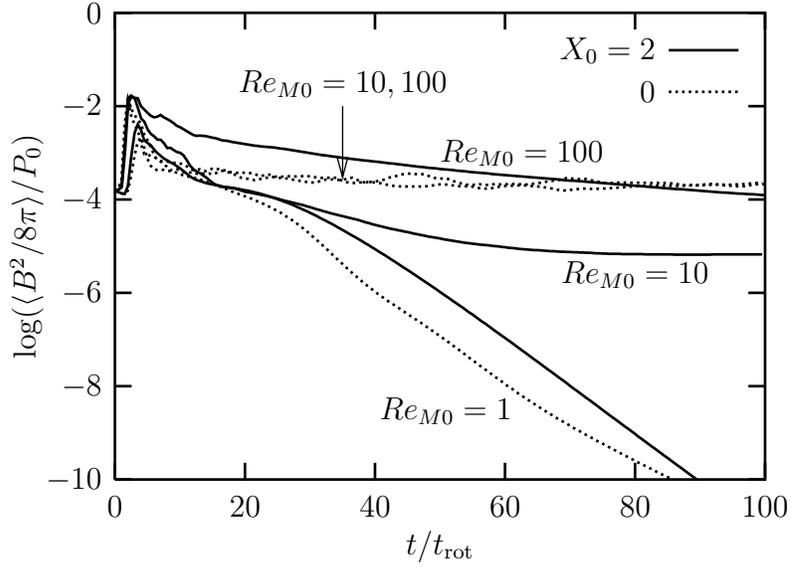

\end{document}